\documentclass[fleqn,usenatbib]{mnras}

\usepackage{newtxtext,newtxmath}
\usepackage[T1]{fontenc}
\usepackage{physics}
\usepackage{amsmath}
\usepackage[flushleft]{threeparttable}

\DeclareRobustCommand{\VAN}[3]{#2}
\let\VANthebibliography\thebibliography
\def\thebibliography{\DeclareRobustCommand{\VAN}[3]{##3}\VANthebibliography}

\usepackage{graphicx}	% Including figure files
\usepackage{amsmath}	% Advanced maths commands
\title[Bond albedo and energy balance of Uranus]{The bolometric Bond albedo and energy balance of Uranus}

\author[P. G. J. Irwin et al.]{Patrick G. J. Irwin,$^{1}$\thanks{E-mail: patrick.irwin@physics.ox.ac.uk (PGJI)}
Daniel D. Wenkert,$^{2}$
Amy A. Simon,$^{3}$
Emma Dahl, $^{4}$
and Heidi B. Hammel $^{5}$
\\
$^{1}$Atmospheric, Oceanic and Planetary Physics, Department of Physics, University of Oxford, Parks Rd, Oxford, OX1 3PU, UK\\
$^{2}$Jet Propulsion Laboratory, California Institute of Technology, 4800 Oak Grove Drive, Pasadena, CA~91109, USA\\
$^{3}$Solar System Exploration Division/690, NASA Goddard Space Flight Center, 8800 Greenbelt Rd, Greenbelt, MA~20771, USA\\
$^{4}$Department of Geological and Planetary Sciences, California Institute of Technology, 1200 E. California Blvd., Pasadena, CA~91125, USA\\
$^{5}$Association of Universities for Research in Astronomy, 1331 Pennsylvania Ave NW, Suite 1475, Washington, DC 20004, USA
}
\date{Accepted XXX. Received YYY; in original form ZZZ}

\pubyear{2025}

\begin{document}
\label{firstpage}
\pagerange{\pageref{firstpage}--\pageref{lastpage}}
\maketitle

\begin{abstract}
Using a newly developed `holistic' atmospheric model of the aerosol structure in Uranus's atmosphere, based upon observations made by HST/STIS, Gemini/NIFS and IRTF/SpeX from 2000 -- 2009, we make a new estimate the bolometric Bond albedo of Uranus during this time of $A^* = 0.338 \pm 0.011$, with a phase integral of $q^* = 1.36 \pm 0.03$.  Then, using a simple seasonal model, developed to be consistent with the disc-integrated blue and green magnitude data from the Lowell Observatory from 1950 -- 2016, we model how Uranus's reflectivity and heat budget vary during its orbit and determine new orbital-mean average values for the bolometric Bond albedo of  $\overline{A^*} = 0.349 \pm 0.016$ and 
for the absorbed solar flux of $\overline{P_\mathrm{in}}=0.604 \pm 0.027$ W m$^{-2}$. Assuming the outgoing thermal flux to be $\overline{P_\mathrm{out}}=0.693 \pm 0.013$ W m$^{-2}$, as previously determined from Voyager 2 observations, we arrive at a new estimate of Uranus's average heat flux budget of $P_\mathrm{out}/P_\mathrm{in} = 1.15 \pm 0.06$, finding considerable variation with time due to Uranus's significant orbital eccentricity of 0.046. This leads the flux budget to vary from $P_\mathrm{out}/P_\mathrm{in} = 1.03$ near perihelion, to 1.24 near aphelion.  We conclude that although $P_\mathrm{out}/P_\mathrm{in}$ is considerably smaller than for the other giant planets, Uranus is not in thermal equilibrium with the Sun. 
\end{abstract}

\begin{keywords}
planets and satellites: atmospheres -- radiative transfer -- scattering -- techniques: photometric
\end{keywords}

%%%%%%%%%%%%%%%%%%%%%%%%%%%%%%%%%%%%%%%%%%%%%%%%%%

%%%%%%%%%%%%%%%%% BODY OF PAPER %%%%%%%%%%%%%%%%%%

\section{Introduction}

All the solar system giant planets appear to emit more thermal radiation to space than they absorb from the Sun, indicating that they have substantial reservoirs of internal heat left over from formation. For planets in thermal equilibrium with the Sun the ratio of the total power thermally emitted to the total solar power absorbed, i.e.,  $P_\mathrm{out}/P_\mathrm{in}$, would be equal to 1.0, but generally accepted estimates of this ratio for the giant planets, based on observations made by the NASA Voyager missions,  are Jupiter: $1.67 \pm 0.09$ \citep{hanel81}, Saturn: $1.78 \pm 0.09$ \citep{hanel83}, Uranus: $1.06 \pm 0.08$ \citep{pearl90}, and Neptune: $2.61 \pm 0.28$ \citep{pearl91}, although  the values for Jupiter and Saturn have recently been updated from Cassini observations to $2.132 \pm 0.051$ \citep{li18} and $2.39 \pm 0.21$ \citep{wang24}, respectively. The flux ratio seems to be particularly low for Uranus and is consistent with Uranus being in thermal equilibrium with the Sun, which would make it a very notable outlier from the rest of the giant planets. 

The low level of the $P_\mathrm{out}/P_\mathrm{in}$ ratio for Uranus, combined with its very high obliquity (i.e., inclination of its spin axis relative to its orbital plane) of $97^\circ$ led \citet{stevenson86} to suggest that towards the end of its initial formation Uranus suffered an oblique impact with a large planetesimal (of approximately 2 Earth masses) that tipped Uranus's spin axis over on its side, heated just the outer shell of the planet, which has since cooled, and threw off of a disc of material, which has since coalesced to form a regular, compact satellite system.  \citet{stevenson86} argued that Neptune, in contrast, may have suffered a more head-on collision that created high internal temperatures and did not throw off a disc of material, which could explain Neptune's high observed $P_\mathrm{out}/P_\mathrm{in}$ ratio and absence of a compact satellite system. This hypothesis has been returned to a number of times since, most recently by \citet{reinhardt20}, who performed Smoothed Particle Hydrodynamics (SPH) simulations that support this scenario. Alternatively, it has been proposed that compositional gradients deep in Uranus’s atmosphere may give rise to a deep thermal boundary layer, inhibiting convection and trapping heat \citep{nettelmann16,vazan20, scheibe21}, thus reducing $P_\mathrm{out}$. Another possibility is that since existing observations of Uranus cover only a fraction of a Uranian year, they may have been made during a quiescent meteorological period when heat release was low. In Saturn's atmosphere, for example, increased storm activity was found to affect the planet’s heat budget by modifying both the amount of regional emitted heat and absorbed solar energy \citep{li15}. Increased storm activity at Uranus could result in changes in heat release by varying the atmosphere’s reflectivity and therefore the amount of absorbed sunlight \citep[e.g., ][]{fletcher20}, and also the degree of vertical heat transport via cumulus convection driven by the condensation of species such as CH$_4$, H$_2$O, NH$_4$SH, NH$_3$, and/or H$_2$S \citep{sromovsky05, hueso20}. Finally, there remains the possibility that the Voyager measurement of Uranus's  outgoing thermal flux may be underestimated due to calibration uncertainties \citep[e.g.,][]{li18}. However,  all these conclusions and interpretations are contingent upon the accuracy of the $P_\mathrm{out}/P_\mathrm{in}$  estimates, and it can be seen that for Uranus this ratio is only currently constrained to lie in rather a broad range from 0.98 to 1.14. 

In this paper we use the `holistic' aerosol model of Uranus and Neptune, developed by \citet{irwin22} and based upon observations made by HST\footnote{HST - Hubble Space Telescope}/STIS, Gemini/NIFS and IRTF\footnote{IRTF - NASA Infrared Telescope Facility}/SpeX from 2000 -- 2009 to improve upon the $P_\mathrm{out}/P_\mathrm{in}$ estimate for Uranus. We first determine the bolometric Bond albedo and $P_\mathrm{out}/P_\mathrm{in}$ flux ratio of Uranus in the early 2000s and then use a simple seasonal model, developed by \citet{irwin24} and consistent with the disc-integrated blue and green magnitude data from the Lowell Observatory \citep{lockwood19}, to model how Uranus's reflectivity and heat budget vary during Uranus's year.% We thus determine a new orbital-mean average value for the heat flux budget of $P_\mathrm{out}/P_\mathrm{in} = 1.15 \pm 0.06$, finding considerable variation of this through Uranus's eccentric orbit. We conclude that Uranus is not in thermal equilibrium with the Sun, but is still radiating significant internal heat of formation away into space. However, our determined flux ratio for Uranus is still much smaller than the other three giant planets and is thus still consistent with the relative giant impact formation pathways proposed for Uranus and Neptune.

\begin{table*}
\caption{Determinations of albedo and energy balance of Uranus (Extended from Table 1 of \citet{pearl90}).}
\begin{threeparttable}
\begin{tabular}{l l l l l l l}
\hline
Reference & year & $p$ & $q$ & $A$ & $T_\mathrm{eff}$ & $P_\mathrm{out}/P_\mathrm{in}$ \\
\hline
\citet{lockwood83}\citep[from ][]{younkin70} & 1962 & $0.228 \pm 0.008$ & $1.50^{+0.14}_{-0.17}$ & $0.342 \pm 0.041$ & $57.0 \pm 2.5^\mathrm{a}$ & $0.92 \pm 0.17$ \\
\citet{lockwood83} & 1981 & $0.262 \pm 0.008$ & " & $0.393 \pm 0.041$ & " & $1.00 \pm 0.19$ \\
\citet{neff85} & 1981 & $0.28 \pm 0.02$ & $1.2$ & $0.34 \pm 0.02$ & $58.6 \pm 2.0^\mathrm{b}$ & $1.12 \pm 0.22$ \\
\citet{pollack86} & 1982 -- 1985 & $0.270 \pm 0.020$ & $1.26\pm 0.11$ & $0.343 \pm 0.055$ & $57.7 \pm 2.0^\mathrm{c}$ & $1.10 \pm 0.22$  \\
\citet{pollack86} & orbital mean & $0.253 \pm 0.046$ & " & $0.319 \pm 0.051$ & " & $1.06 \pm 0.21$ \\
\citet{pearl90} & 1986 & $0.231 \pm 0.048$ & $1.40\pm 0.14$ & $0.322 \pm 0.049$ & $59.1 \pm 0.3$ & $1.14 \pm 0.09$ \\
\citet{pearl90} & orbital mean & $0.215 \pm 0.046$ & " & $0.300 \pm 0.049$ & " & $1.06 \pm 0.08$ \\
This investigation & 2000 -- 2009 & $0.249 \pm 0.007$ & $1.36\pm 0.03$ & $0.338 \pm 0.011$ & $59.1 \pm 0.3$ & $1.24 \pm 0.06$ \\
This investigation & orbital mean & $0.257 \pm 0.007$ & " & $0.349 \pm 0.016$ & " & $1.15 \pm 0.06$ \\
\hline
\label{tab:albedos}
\end{tabular}
%   \begin{tablenotes}
      \small
      $^\mathrm{a}$\citet{courtin79}, $^\mathrm{b}$\citet{hildebrand85}, $^\mathrm{c}$\citet{orton85}.
%    \end{tablenotes}
\end{threeparttable}
\end{table*}

\section{Measurements of the Bond Albedo and power balance of Uranus}

At a particular wavelength the geometric albedo of a planet, $p$, is defined as the power of the reflected sunlight at a phase angle of $0^\circ$ compared with that from  a flat, perfectly reflecting Lambertian surface of the same cross-sectional area. The Bond albedo $A$, however, is the total fraction of sunlight incident on the planet that is reflected away in \textbf{all} directions and is related to the geometric albedo, $p$, by the relation

\begin{equation}
 A=pq,
 \label{eq:bond}
\end{equation}

where $q$ is the \textit{phase integral}, defined as the integral of the flux $I(\alpha)$ scattered from the planet into phase angle $\alpha$:

\begin{equation}
 q=2\int_0^\pi \frac{I(\alpha)}{I(0)}\sin \alpha \mathrm{d}\alpha.
 \label{eq:phase_int}
\end{equation}

If we integrate the incident and reflected powers over all wavelengths, then we arrive at `integrated' or `bolometric' values of these parameters, which we denote as: $p^*$, $q^*$, and $A^*$.

%The low value of $P_\mathrm{out}/P_\mathrm{in} = 1.06 \pm 0.08$ arises mostly from the low bolometric Bond albedo determined from Voyager IRIS observations by \citet{pearl90} of $\Bar{A} = 0.300 \pm 0.049$. 

The Bond albedo and radiative heat balance of Uranus has been addressed by a number of studies. 
\citet{lockwood83} analysed high resolution spectrophotometry observations of Uranus made in 1981 at the Lowell Observatory and found $p^* = 0.262 \pm 0.008$,  $q^* = 1.50^{+0.14}_{-0.17}$, and  $A^* = 0.393 \pm 0.041$. It was also found that the albedo had risen significantly since previous observations made between 1961 and 1963 \citep{younkin70}, for which the following values were computed: $p^* = 0.228 \pm 0.008$, and  $A^* = 0.342 \pm 0.041$. Using $A^* = 0.393 \pm 0.041$ the estimated absorbed solar power per unit area over the whole planet in 1981 was estimated to be 0.563 -- 0.643 W m$^{-2}$, while assuming the effective black body temperature of Uranus in the infrared to be $T_\mathrm{eff}=57.0 \pm 2.5$K \citep{courtin79}, the outgoing flux was estimated to be 0.500 -- 0.711 W m$^{-2}$. Hence, \citet{lockwood83} estimated the internal heat flux to be $< 0.122$ W m$^{-2}$. From these results (summarised in Table \ref{tab:albedos}), we here derive an estimate (adding errors in quadrature) of  $P_\mathrm{out}/P_\mathrm{in} = 1.00 \pm 0.19$. For the 1961 - 1963 data, we estimate from these data that $P_\mathrm{out}/P_\mathrm{in} = 0.92 \pm 0.17$

\citet{neff85} analysed high resolution spectrophotometry observations of Titan, Uranus and Neptune made in around 1981 by the International Ultraviolet Explorer \citep{caldwell81}, Macdonald Observatory \citep{neff84} and IRTF, and estimated $p^* = 0.28 \pm 0.02$,  $q^* = 1.2$, and  $A^* = 0.34 \pm 0.02$. The absorbed solar flux was estimated to be the same as that emitted from a black body of temperature $T=57.0 \pm 2.0$K. Assuming the effective black body temperature of Uranus in the infrared  to be $T_\mathrm{eff}=58.6 \pm 2.0$K \citep{hildebrand85}, the internal luminosity of Uranus was estimated to be $L_\mathrm{int}=(0.6 \pm 1.4) \times 10^{15}$ W. From these results we estimate the internal heat flux (using their assumed Uranus radius of 25,250 km) to be  $\le 0.25$ W m$^{-2}$, and from this we here derive an estimate (adding errors in quadrature)  of $P_\mathrm{out}/P_\mathrm{in} = 1.12 \pm 0.22$ (Table \ref{tab:albedos}).

\citet{pollack86} constructed an atmospheric aerosol model that was consistent with phase angle observations (0 -- $85^\circ$) from ground-based observatories \citep{neff84} and Voyager-1/ISS observations made between 1982 and 1985. From this model,  \citet{pollack86} calculated that during this period, $p^* = 0.27 \pm 0.02$,  $q^* = 1.26 \pm 0.11$, and  $A^* = 0.343 \pm 0.055$. Since \citet{lockwood83} noted a 14\% increase in Uranus's albedo from 1961--1963 to 1981, \citet{pollack86} assumed that the orbitally-averaged values of their albedos should be reduced by 7\% to $\overline{p^*} = 0.253 \pm 0.046$, and  $\overline{A^*} = 0.319 \pm 0.051$.  The orbitally-averaged absorbed solar flux was estimated to be the same as the flux emitted from a black body of temperature $T=56.9 \pm 2.0$K. Assuming the effective black body temperature of Uranus in the infrared  to be  $T_\mathrm{eff}=57.7 \pm 2.0$K \citep{orton85}, \citet{pollack86} concluded that the internal heat flux was $\le 0.27$ W m$^{-2}$. From these results, we here derive an estimate of $P_\mathrm{out}/P_\mathrm{in} = 1.06 \pm 0.21$ (Table \ref{tab:albedos}). We have here also scaled this value (using the Bond albedo estimates) to an equivalent flux ratio in 1982 -- 1985 of $P_\mathrm{out}/P_\mathrm{in} = 1.10 \pm 0.22$.

\begin{figure*}
	\includegraphics[width=\textwidth]{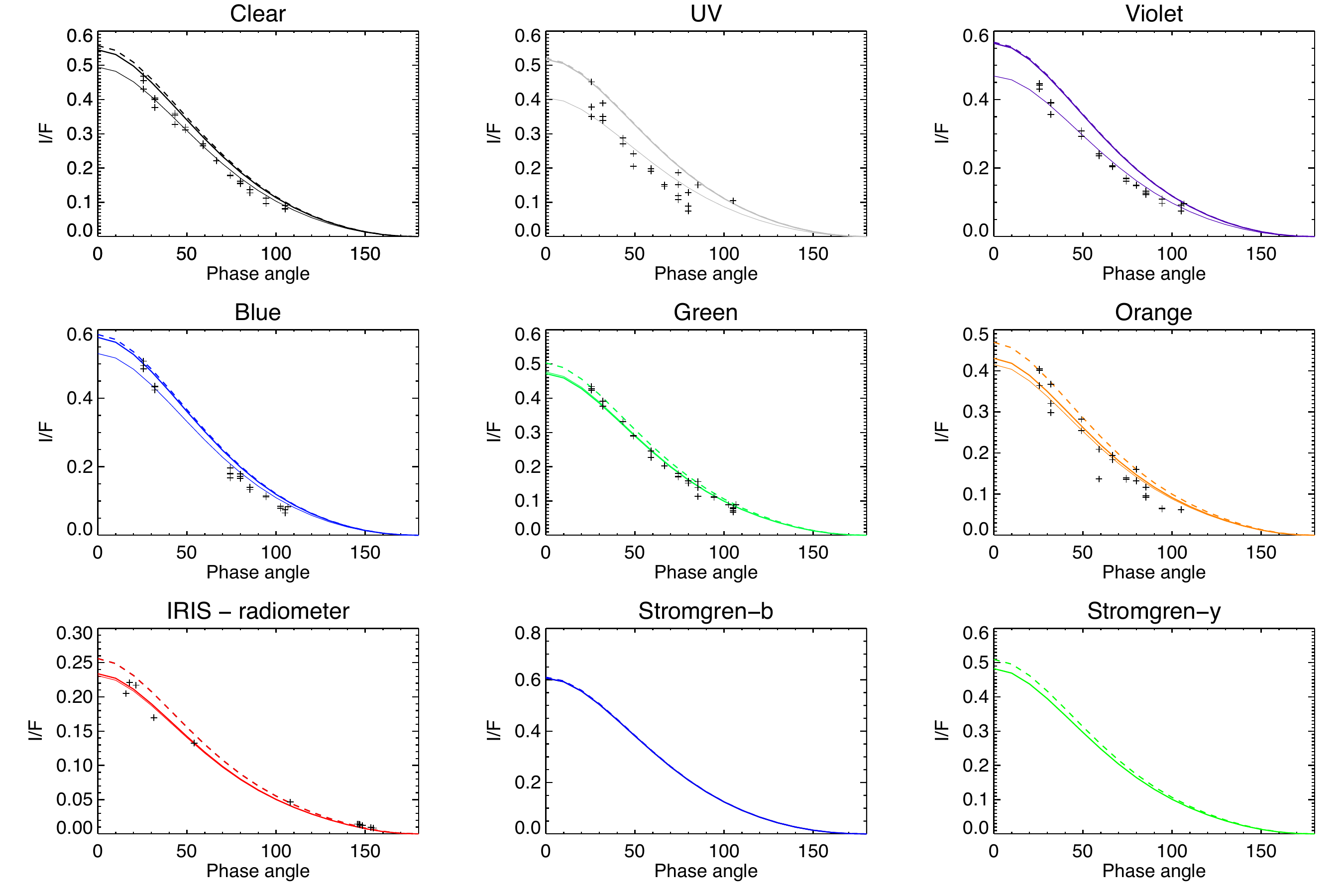}
    \caption{Phase curves calculated from our NEMESIS simulations of Uranus and averaged over different filter profiles of Voyager/ISS, Voyager/IRIS and, for reference, Str\"omgren-$b$ and Str\"omgren-$y$. The Voyager phase curve simulations are compared with the ISS observations processed by \citet{wenkert23} and IRIS observations processed by \citet{pearl90}. Here the solid lines show the calculations from our model atmosphere fitted to observations between 2000 and 2009, while the dashed lines show calculations that have been corrected to the time of the Voyager-2 observations in 1986, when Uranus was more reflective,  using the seasonal albedo curves shown later in Fig. \ref{fig:seasonal_meas_sim}. The thin lines are the simulated curves scaled to match the observations most closely, showing the excellent modelling of the shape of these curves, if not necessarily the exact photometric calibration for each filter. }
    \label{fig:phase_curves}
\end{figure*}

\citet{pearl90} used Voyager/IRIS-radiometer observations of Uranus made during the Voyager 2 flyby of Uranus in January 1986 to estimate Uranus's Bond albedo. Measurements of the reflectivity of Uranus, integrated over the IRIS radiometer, were made at several phase angles and are shown in Fig. \ref{fig:phase_curves} and compared with the Voyager-1/ISS phase angle observations used by \citet{pollack86}, and more recently extended by \cite{wenkert23}. The IRIS radiometer is sensitive from 0.3 -- 1.9 $\mu$m and the IRIS radiometer sensitivity  function is compared with the Voyager/ISS filter functions and a reference Uranus spectrum in Fig. \ref{fig:spectra}. \citet{pearl90} assumed that the IRIS-radiometer-averaged reflectivities were equivalent with the bolometric values and thus reported $p^*=0.231 \pm 0.048$, $q^* = 1.4 \pm 0.14$, and $A^*=0.322 \pm 0.049$.  These observations were made in 1986 near Uranus's solstice, when Uranus is at its brightest. \citet{pearl90} noted that their albedoes were thus likely to be greater than the average value and so revised their Bond albedo figure down by 7\% \citep[from ][]{lockwood83} to arrive at an estimate of the orbital mean Bond albedo of $\overline{A^*}=0.300 \pm 0.049$. By analyzing the IRIS infrared spectrometer data (covering $\sim$11 to 55 $\mu$m) and extending to longer wavelengths with modelling,  \citet{pearl90} then estimated that the overall infrared flux emitted into all directions to be $F = 0.693 \pm 0.013$ W m$^{-2}$, equivalent to a black body of temperature of $T=59.1 \pm 0.3$K, which is larger than all the other equivalent black body temperatures assumed in the previous studies discussed here. Using these numbers \citet{pearl90} concluded that $P_\mathrm{out}/P_\mathrm{in} = 1.06 \pm 0.08$.  We have here also scaled this (from the Bond albedoes) to an equivalent flux ratio in 1986 of $P_\mathrm{out}/P_\mathrm{in} = 1.14 \pm 0.09$.

The comparison of these estimates for Uranus's albedo and radiative heat balance from these studies is summarised in Table \ref{tab:albedos}.

\section{Simulating the Bond Albedo of Uranus} \label{sec:2}

In this study we simulate the reflectivity spectrum of Uranus using our NEMESIS \citep{irwin08} radiative transfer and retrieval model. NEMESIS uses the plane-parallel doubling and adding multiple scattering scheme of \citet{plass73} and has been used in several radiative transfer studies of Uranus and Neptune.  Using NEMESIS, \citet{irwin22} recently developed an `holistic' aerosol model for Uranus and Neptune, which was simultaneously consistent with HST/STIS spectral imaging observations from 0.30 -- 1.02 $\mu$m, Gemini/NIFS spectral imaging observations from 1.4 -- 1.8 $\mu$m, and a central meridian line-averaged IRTF/SpeX observation from 0.8 -- 1.9 $\mu$m. For Uranus, the HST/STIS observations  were made in 2002 \citep{kark09}, the Gemini/NIFS observations made in 2009 \citep{irwin11}, and the IRTF/SpeX observation  made in 2000 \citep{rayner09}.  This model consisted of a vertically-extended semi-infinite deep cloud deck `Aerosol-1', a vertically-thin middle `Aerosol-2' layer, centred at 1--2 bar, coincident with the methane condensation level, and a vertically-extended upper-tropospheric `Aerosol-3' haze at pressures less than 1.5 bar. By adjusting the cloud opacities, particle sizes, complex refractive index spectra of these three layers and also varying the pressure of the Aerosol-2 layer, the model was able to reproduce the observed reflectivity and limb-darkening curves of both Uranus and Neptune, where the limb-darkening was found to be consistent with the model of \citet{minnaert41}:
\begin{equation}
(I/F) = (I/F)_{0}\mu_{0}^{k} \mu^{k-1}.
\label{eq:minnaert}
\end{equation}

Here,  $\mu$ is the cosine of the viewing zenith angle, $\mu_0$ is the cosine of the solar zenith angle and the fitted parameters are the nadir reflectances $(I/F)_0$ and the limb-darkening coefficients $k$.

This holistic model has since been slightly revised by \citet{james23} and \citet{irwin24}, with the semi-infinite Aerosol-1 component replaced by a vertically-thin cloud deck centred at $\sim$5 bar. This revised model was found to fit the observations just as well, but made it easier to disentangle the spectral effects of the Aerosol-1 and Aerosol-2 layers, leading to more reliable and robust retrievals.

The disc-integrated magnitude of Uranus at Str\"omgren-$y$ and -$b$ wavelengths was measured nearly annually from 1950 to 2016 by observers at the Lowell Observatory \citep{lockwood19}, who compiled a unique record of the seasonal variation of Uranus's brightness and colour. \citet{irwin24} converted these magnitude data to disc-averaged reflectivities and compared these with HST/WFC3 observations from 2014 -- 2022, and HST/STIS observations made in 2002, 2012 and 2015 \citep{kark09, sromovsky14, sromovsky19}, finding very good photometric agreement between all three datasets. %Hence, we believe the Lowell reflectivities to be reliable and have used these as the standard reference in this study. 
\citet{irwin24} then used the revised holistic model to simulate the seasonal variation in Uranus's albedo at Str\"omgren-$y$ and -$b$ wavelengths.

Taking our best modified holistic model fit to the Uranus data of \citet{irwin22}  we computed the mean reflectivity spectrum of Uranus's atmosphere from 0.2 - 1.9 $\mu$m at all possible solar and viewing angles  using NEMESIS \citep{irwin08}, where for these simulations we used 21 angles in our zenith angle quadrature scheme and 38 Fourier components in the azimuthal decomposition. The computed spectra were then used to simulate the overall reflectivity spectrum of Uranus for different phase angles by interpolating the spectra to the local viewing zenith angle, solar zenith angle and azimuth angle at each location on the planet and integrating over the disc. The reflectance spectra were converted to disc-integrated spectral irradiance (W m$^{-2}$ $\mu$m$^{-1}$) by multiplying by our standard reference solar spectrum, which was taken to be that of \citet{chance10}, 
extrapolated to longer/shorter
wavelengths with the solar spectrum of \citet{kurucz93}. To simulate the visible appearance of Uranus at different phase angles, the interpolated spectra at each location on the planet were convolved with standard XYZ colour-matching functions and then converted to the sRGB colour space as described by \citet{stockman00}, \citet{stockman19} and \citet{irwin24}. In Fig. \ref{fig:colour_images} we show the simulated `true-colour' appearance of Uranus at phase angles 0$^\circ$, 30$^\circ$, 60$^\circ$, 90$^\circ$, 120$^\circ$ and 150$^\circ$, where the images have been Gamma-corrected to give the appearance of Uranus as would be seen by an average human observer \citep[see ][]{irwin24}. As can be seen the apparent colour of Uranus becomes less blue as we go to higher phase angles and the integrated reflectivity drops\footnote{This best-fit model was also used to generate a revised simulated video of a future possible spacecraft's approach to Uranus and the descent of an entry probe (\url{https://www.youtube.com/watch?v=NPZeBLCjI-0}), where this colour variation with phase angle can also be seen.}.

\begin{figure*}
	\includegraphics[width=\textwidth]{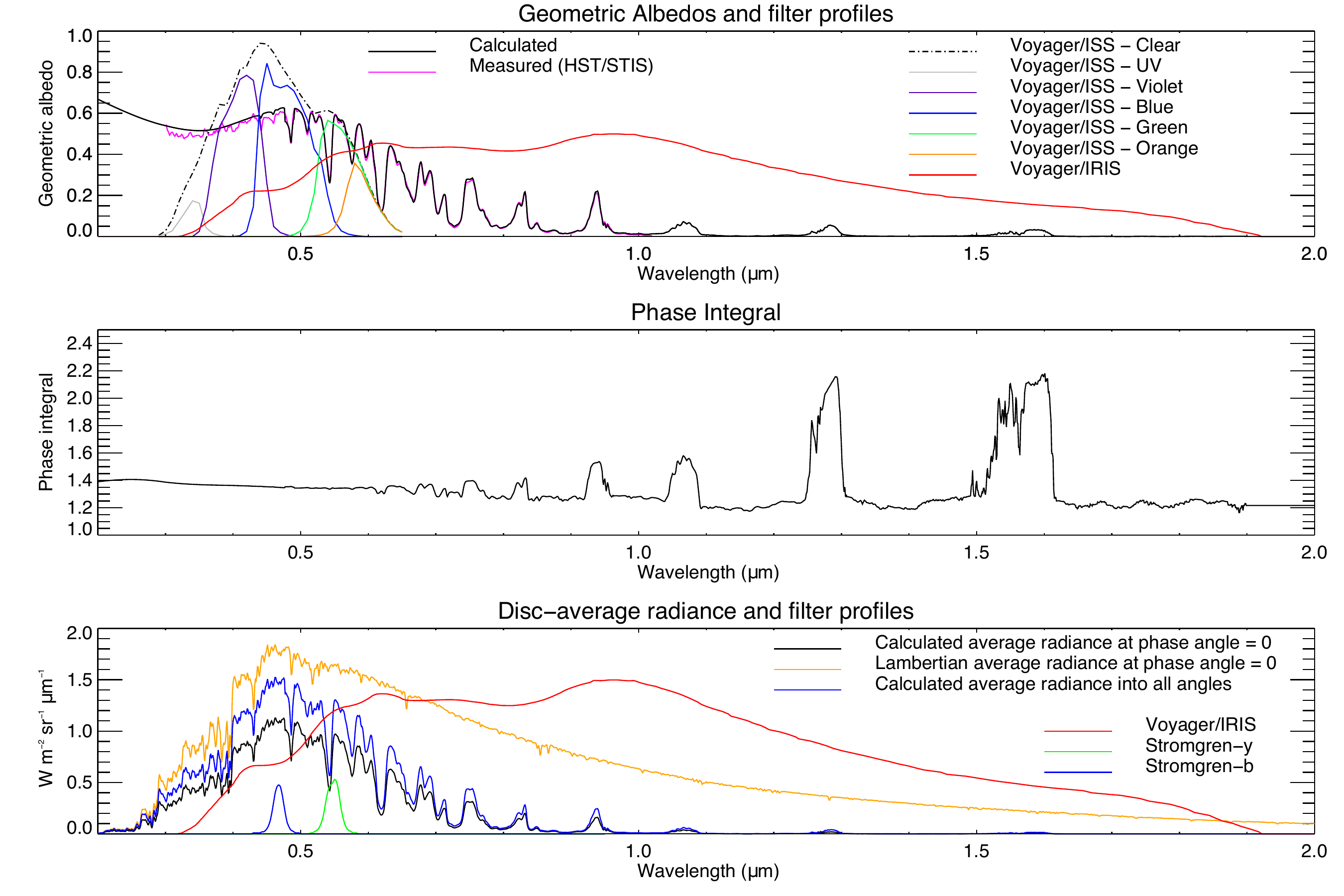}
    \caption{Simulated and measured spectra of Uranus. Top plot compares the simulated geometric albedo spectrum (black line) compared with the Voyager/ISS filter functions \citep{danielson81} and the Voyager/IRIS radiometer response function (red) \citep{hanel81}. In addition, we show the measured HST/STIS albedo spectrum of \citet{kark09} (pink line), showing that our model agrees well with the geometric albedo spectrum measured in 2002 by HST/STIS, as expected. Middle panel shows our calculated phase integral spectrum $q(\lambda)$ from these calculations. Bottom plot compares the modelled disc-averaged radiance spectrum (calculated at phase angle $\alpha=0^\circ$,  black) with the expected radiance from a perfectly-scattering Lambertian surface (orange). In addition, we show the disc-averaged reflectance into all angles (blue) and compare this with the Str\"omgren-$b$ and Str\"omgren-$y$, and Voyager/IRIS radiometer filter functions (red).}
    \label{fig:spectra}
\end{figure*}

The computed disc-integrated spectra at each phase angle were then convolved with the Voyager/ISS filter functions and also the Voyager/IRIS radiometer responsivity function to calculate filter-averaged reflectivities of Uranus as a function of phase angle. These phase angle calculations are compared in Fig. \ref{fig:phase_curves} with the Voyager/ISS observations of \citet{wenkert23a} (from which phase curves have been extracted \citep{wenkert23}) and the Voyager/IRIS observations of \citet{pearl90}. In addition, simulated phase curves at Str\"omgren-$b$ and -$y$ wavelengths are also shown. It should be noted that Voyager observed increasingly equatorial regions of Uranus as it headed to higher phase angles due to the large inclination of Uranus’s axis and the fact that these observations were made near solstice. Since Uranus is known to form a `hood' of increased haze opacity at polar latitudes near solstice, this means that the reflectivities observed by Voyager at different phase angles are weighted to regions of very different aerosol structure, which adds to the complexity of interpreting these phase curves. 

It can be seen that the observed variation of reflectivity with phase angle is reasonably well defined in the Clear, Violet, Blue, and Green channels, where the signal-to-noise ratio of the Voyager/ISS observations are good, but is less clear in the UV and Orange channels. The UV and Orange reflectivities are known to be the least reliable observations in this dataset, due to increased noise caused by the lower radiances seen in these channels (Fig. \ref{fig:spectra}) and insufficient exposure times. However, it can be seen that the simulated phase curves are in reasonably good agreement with observations, which adds confidence that NEMESIS models the reflectivity of Uranus at different phase angles reliably. Indeed, if we scale our simulations by single factor for each filter (for example for the Clear filter a factor of 0.908 has been applied), it can be seen that the simulations are in good agreement with the shape of the measured phase curves, which suggests that the differences could just be caused by photometric inconsistencies; such channel-to-channel variation was also seen when we analysed Voyager-2 ISS observations of Uranus \citep{irwin24}.   We thus used our NEMESIS simulations to compute the phase integral $q$ (Eq. \ref{eq:phase_int}) numerically, validating these calculations against a model planet with a uniform perfectly scattering Lambertian surface, for which we computed the expected values of $p=2/3$ and $q=1.5$, leading to $A=1$.

\begin{figure}
	\includegraphics[width=\columnwidth]{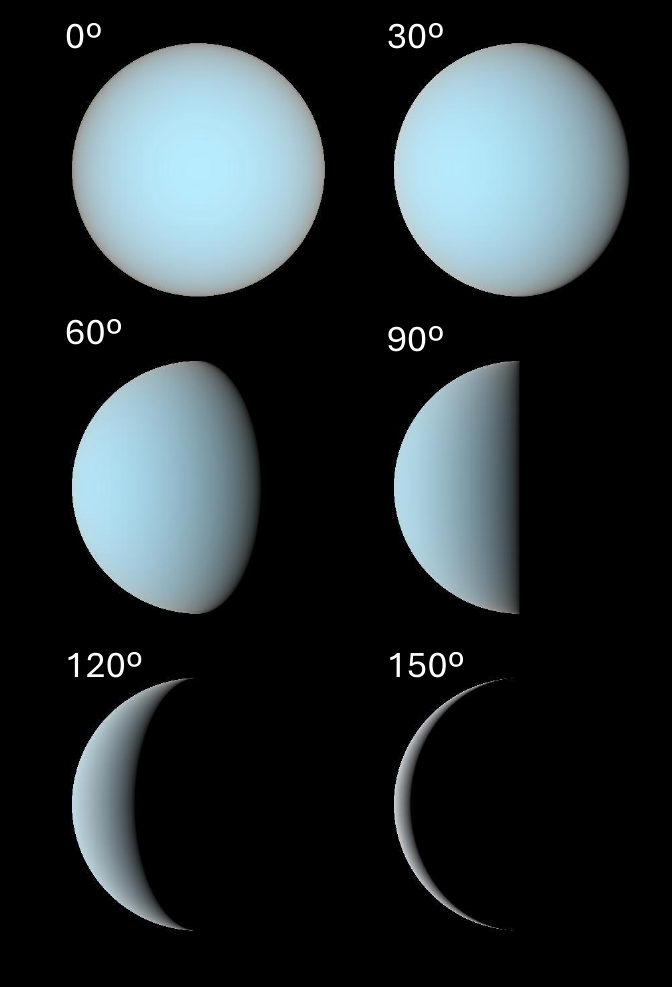}
    \caption{Computed `true-colour' appearance of Uranus calculated with our NEMESIS radiative transfer code  at six different representative phase angles. Following the procedure of \citet{irwin24}, the computed spectra in each pixel have been converted to the XYZ colour system and then to sRGB, before finally being  Gamma-corrected to best represent the appearance of Uranus to the average human observer. The same intensity scale has been used for all figures in this composite image.}
    \label{fig:colour_images}
\end{figure}

The calculated geometric albedo spectrum $p(\lambda)$ from our model is shown in Fig. \ref{fig:spectra} and can be seen to be in good agreement with the measured disc-averaged HST/STIS albedo spectrum of \citet{kark09}, from which this model was derived. Figure \ref{fig:spectra} also shows the Voyager/ISS channel filter profiles \citep{danielson81}, the Voyager/IRIS-radiometer spectral response functions \citep{hanel81}, and the Str\"omgren-$b$ and Str\"omgren-$y$ filter functions \citep{crawford70}. As can be seen, the observed and  modelled geometric albedo spectrum of Uranus varies greatly with wavelength, with high albedos seen at visible and UV wavelengths, but decreasing rapidly at longer wavelengths due to the increasing absorption of gaseous methane. Hence, the mean albedo of a filter passband depends strongly on these responsivity curves.

From the computed irradiance spectra $I(\lambda, \alpha)$ (W m$^{-2}$ $\mu$m$^{-1}$) at wavelength $\lambda$ and phase angle $\alpha$ we calculated the phase integral spectrum $q(\lambda)$, which is also shown in Fig. \ref{fig:spectra}. It can be seen that our computed phase integrals lie in the range 1.3 to 1.4 at visible wavelengths, reducing to 1.2 to 1.3 at longer methane-absorbing wavelengths, but increasing in the methane windows. This is expected as the main reflection is from the Aerosol-`2' layer at $\sim$1.5 bar, which is modelled as being  composed of micron-sized particles that are rather forward scattering at these wavelengths, thus increasing $q$. We calculated filter-averaged geometric and Bond albedoes using the following equations:%Eqs. \ref{eq:albedo}. 

\begin{equation}
 \Bar{p}=\frac{\int f(\lambda)I(\lambda, 0) \mathrm{d}\lambda}{\int f(\lambda) S(\lambda) \mathrm{d}\lambda}, \quad \mathrm{and} \quad \Bar{A}=\frac{\int f(\lambda)q(\lambda)I(\lambda, 0) \mathrm{d}\lambda}{\int f(\lambda) S(\lambda) \mathrm{d}\lambda}.
 \label{eq:albedo}
\end{equation}

%and

%\begin{equation}
% \Bar{A}=\frac{\int f(\lambda)q(\lambda)R(\lambda, 0) \mathrm{d}\lambda}{\int f(\lambda) S(\lambda) \mathrm{d}\lambda}
% \label{eq:bond_alb}
%\end{equation}

Here, $S(\lambda)$ is the solar irradiance spectrum for a perfectly reflecting Lambertian disc of Uranus's size and distance from the Sun, and $f(\lambda)$ is the filter sensitivity profile. It is important to stress that we computed these filter-averaged albedos by taking ratios of the filter-averaged irradiances, rather than filter-averaging the albedo spectra $p(\lambda)=I(\lambda,0)/S(\lambda)$ and $A(\lambda)=q(\lambda)I(\lambda,0)/S(\lambda)$ using:% Eqs. \ref{eq:alb_wrong}.

\begin{equation}
 \tilde{p}=\frac{\int f(\lambda)p(\lambda) \mathrm{d}\lambda}{\int f(\lambda) \mathrm{d}\lambda}, \quad \mathrm{and} \quad \tilde{A}=\frac{\int f(\lambda)A(\lambda) \mathrm{d}\lambda}{\int f(\lambda) \mathrm{d}\lambda}.
 \label{eq:alb_wrong}
\end{equation}

%and

%\begin{equation}
% \tilde{A}=\frac{\int f(\lambda)A(\lambda) \mathrm{d}\lambda}{\int f(\lambda) \mathrm{d}\lambda}
% \label{eq:bond_alb_wrong}
%\end{equation}

The spectra of both Uranus and the Sun can be seen to vary greatly with wavelength in Fig. \ref{fig:spectra}. Hence, while $\Bar{p} \sim \tilde{p}$ and $\Bar{A} \sim \tilde{A}$ for narrow filters, it is not a good approximation for the wide filter passbands we consider here. 

Our computed averaged albedos for a range of filters and wavelength ranges are listed in Table \ref{tab:calculations}. In this table, the quoted effective phase integrals are computed as $\Bar{q}=\Bar{A}/\Bar{p}$. Since the reflectivity of Uranus decreases greatly with wavelength, the filter-averaged albedoes for the Voyager/IRIS-radiometer filter are significantly lower than those of shorter-wavelength filters. It can also be seen in Fig. \ref{fig:spectra} that the IRIS radiometer filter biases against the reflectivity at very short wavelengths. If instead we calculate the albedoes using a filter that is uniformly transmitting from 0.2 to 1.9 $\mu$m, we find significantly higher albedoes. However, even this calculation ignores some reflected radiance, as the solar spectrum is non-negligible at longer wavelengths and Uranus still has some reflectivity. To address this, we first matched a combined reference IRTF/SpeX and JWST/NIRSpec spectrum to the computed $I(\lambda,\alpha)$ radiance spectra from 1.5 to 1.6 $\mu$m and used these scaled spectra to extend the calculation from 1.9 $\mu$m to 2.5 $\mu$m. Secondly, since $\sim$3.7\% of the Sun's irradiance is at wavelengths longer then 2.5 $\mu$m, we extended the computed spectra further to 100 $\mu$m, assuming that Uranus has insignificant reflectivity at wavelengths longer than 2.5 $\mu$m (i.e., we assumed Uranus's reflectivity at wavelengths longer than 2.5 $\mu$m to be zero). We believe this to be a sensible assumption as Uranus's observed reflectivity is already very small at 2.5 $\mu$m, and aerosol particles would have to be very large and the layers optically thick to have significant backscatter at longer wavelengths, which is not consistent with current atmospheric haze models \citep[e.g., ][]{irwin22}. As can be seen, these extensions have the effect of slightly decreasing the computed albedos, but they are still 6\% greater than the Voyager/IRIS estimate. 
Integrating over all wavelengths (0.2 -- 100 $\mu$m) we conclude that in the period 2000 -- 2009: $p^*=0.249\pm 0.007$, $q^*=1.36 \pm 0.03$, and $A^* = 0.338 \pm 0.011$. Our estimate for $q^*$ is well within the range of previous estimates (Table \ref{tab:albedos}), but as we will see in the next section $p^*$ and $A^*$ vary significantly with time, and this variation must be accurately accounted for in order to arrive at robust estimates of the orbital-mean Bond albedo and radiative balance. 

Averaging our simulated spectra over the Voyager/IRIS-radiometer responsivity function, we calculate values of $\Bar{p}_\mathrm{IRIS} = 0.234$, $\Bar{A}_\mathrm{IRIS} = 0.316$ and $\Bar{q}_\mathrm{IRIS}=1.35$. With regard to possible errors, the 2000 -- 2009 observations that our model is derived from are believed to be accurate to 1 to 2\%, and we are able to fit to these data to within this precision. The error on the calculation of the phase integral is harder to check, but the shape of our phase curves looks consistent to first order with the Voyager phase curves shown in Fig. \ref{fig:phase_curves}. Assuming 2\% for the observation and fitting errors and adding these in quadrature, this amounts to $\pm$2.8\% for $p$, giving $\Bar{p}_\mathrm{IRIS}=0.234\pm 0.007$. Assuming 2\% uncertainty for $q$ gives $\Bar{q}_\mathrm{IRIS}=1.36 \pm 0.03$ and combining these errors in quadrature gives $\Bar{A}_\mathrm{IRIS} = 0.316 \pm 0.011$. Our estimates for these IRIS albedoes from 2000 -- 2009 are thus consistent with the values determined by \citet{pearl90} in 1986 of $\Bar{p} = 0.231 \pm 0.048$ and $\Bar{A}=0.322\pm0.049$. However,  we will see in the next section that the albedo of Uranus is known to have decreased significantly from 1986 to the early 2000s, which leads us to suspect that the albedoes of \citet{pearl90} may be underestimated. This conclusion is also consistent with Table \ref{tab:albedos}, where it can be seen that values of \citet{pearl90} are lower than the previous three studies considered.

\begin{table}
\caption{Simulated filter-averaged Bond and geometric albedoes computed from our reference holistic model fitted to observations from 2000 -- 2009. The likely errors on these values is discussed in the main text.}
\begin{tabular}{l l l l}
\hline
Instrument or range &  Geometric albedo &  Bond albedo & Phase integral  \\
  &   $\Bar{p}$ &   $\Bar{A}$ &  $\Bar{q}$ \\
\hline
Str\"omgren-$b$ (467 nm)    &  0.606742   &  0.819338  &    1.35039 \\
Str\"omgren-$y$ (551 nm)   &   0.482306   &  0.647710   &   1.34295 \\
Voyage/ISS -- Clear   &   0.545037  &   0.737174  &    1.35252 \\
Voyager/ISS -- UV      &   0.517394  &   0.709903  &    1.37207 \\
Voyager/ISS -- Violet  &   0.564171  &   0.767374  &    1.36018 \\
Voyager/ISS -- Blue    &   0.577464  &   0.779424  &    1.34974 \\
Voyager/ISS -- Green 1  &   0.474723  &   0.637848  &    1.34362 \\
Voyager/ISS -- Green 2  &   0.471953  &   0.634109  &    1.34359 \\
Voyager/ISS -- Orange  &   0.430355  &   0.578334  &    1.34385 \\
HST/STIS (0.3 -- 1.02 $\mu$m)  &  0.344325   &  0.465665   &   1.35240 \\
Voyager/IRIS Radiometer    &  0.233810   &  0.316006   &   1.35155 \\
0.2 -- 1.9 $\mu$m  &   0.268850  &   0.364617 &     1.35621 \\
0.2 -- 2.5 $\mu$m   &  0.258426   &  0.350479  &    1.35620 \\
0.2 -- 100 $\mu$m   &  0.248886  &   0.337540  &    1.35620 \\
\hline
\label{tab:calculations}
\end{tabular}
\end{table}

\section{Seasonal Cycle Modelling}

\begin{figure*}
	\includegraphics[width=\textwidth]{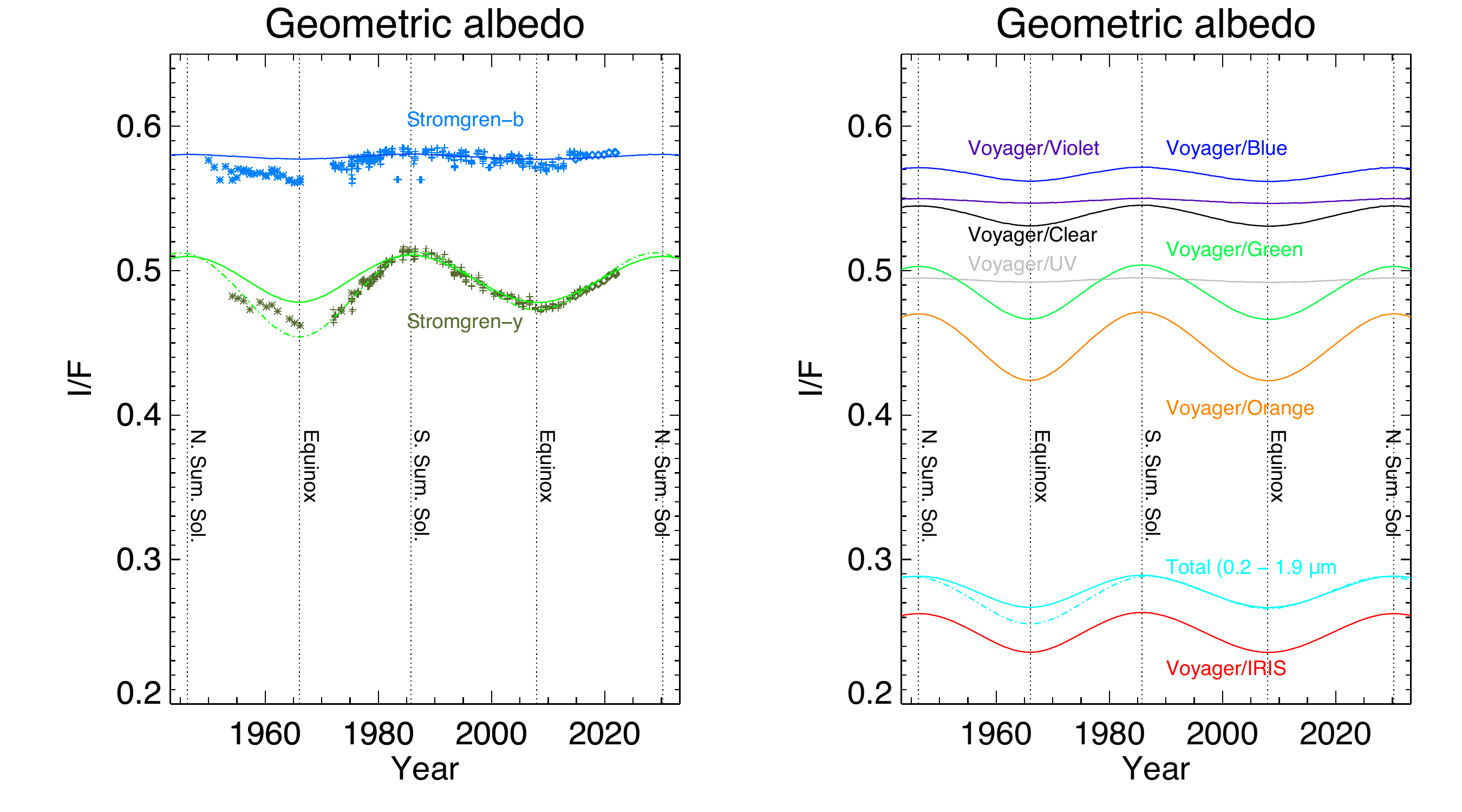}
    \caption{Seasonal variation in reflectivity of Uranus averaged over different filter profiles using a modified seasonal cycle model of \citet{irwin24}, which has an additional component ($\tau = 5$ at 800 nm) of conservatively-scattering particles of mean radius 0.8 $\mu$m added to the Aerosol-2 layer at 1.4 bar at polar latitudes. Left: The calculated reflectivity curves (solid lines) averaged over the Str\"omgren-$b$ and Str\"omgren-$y$ filters are compared with the measured disc-averaged reflectivities calculated by \citet{irwin24} from the magnitude observations of \citet{lockwood19}, recorded directly in these filters from 1972 - 2016 (crosses) or converted from Johnson B and V observations from 1950 - 1966 (asterisks),
    and equivalent HST/WFC3 observations from 2016 - 2022 (diamonds). The dash-dot green line shows an empirically constructed albedo curve that better matches the observed Str\"omgren-$y$ seasonal curve. Right: the calculated reflectivity curves averaged over the different Voyager/ISS filters are shown, together with the curves averaged over the Voyager/IRIS-radiometer sensitivity function, or over all computed wavelengths (0.2 -- 1.9$\mu$m). For this last calculation, the dash-dot cyan line shows an alternative total curve scaled from the empirically fitted Str\"omgren-$y$ seasonal curve.}
    \label{fig:seasonal_meas_sim}
\end{figure*}

Our Bond albedo estimate comes from our modified holistic model fit to HST/STIS, IRTF/SpeX and Gemini/NIFS observations made from 2000 -- 2009 and near equinox when the albedo of Uranus is known to be significantly lower than it is near solstice \citep{lockwood19}. To compare more directly with Voyager-2 observations we need to correct for this albedo change. Furthermore, in order to best estimate the orbital-mean Bond albedo, we need to accurately model the average absorbed solar energy during Uranus's orbit about the Sun.

\citet{irwin24} found that the seasonal variations in reflectivity determined from the Lowell Observatory observations of \citet{lockwood19} could be moderately well approximated by a model where polewards of  40$^\circ$S and 40$^\circ$N an additional opacity ($\tau=5$ at 800 nm) of conservatively-scattering micron-sized particles were added to the Aerosol-2 layer at $\sim$1.5 bar to simulate the effects of Uranus's polar `hood'. Although this was used by \citet{irwin24} to model just Str\"omgren-$b$ and Str\"omgren-$y$ reflectivity variations, we assumed the same model could be used here to simulate the variations in reflectivity at all other wavelengths.

To make the new simulation, we took our reference modified holistic model and computed the Minnaert nadir and limb-darkening spectra $I_0(\lambda)$ and $k(\lambda)$. Then, adding an additional opacity ($\tau = 5$ at 800 nm) of conservatively scattering particles to the Aerosol-2 layer at 1.4 bar with the same mean radius of 0.8 $\mu$m we computed new Minnaert spectra $I_\mathrm{0Pole}(\lambda)$ and $k_\mathrm{Pole}(\lambda)$, more appropriate for polar regions. Then, for one orbit about the Sun, and using geometry ephemeris data from JPL Horizons\footnote{\url{https://ssd.jpl.nasa.gov/horizons/}}, following the procedure of \citet{irwin24}, we computed the radiance at each point on Uranus's disc (including Uranus's polar flattening, which \citet{irwin24} found to have a small effect on the overall reflectivity) as seen from Earth using the Minnaert model for both cases, combining them depending on latitude as:

\begin{equation}
    (I/F) = (1-f)(I/F)_\mathrm{Ref}+f(I/F)_\mathrm{Pole},
\end{equation}
where the weighting factor $f$ was assumed to vary with latitude $\phi$ (in degrees) as:
\begin{equation}
    f=(1+\tanh(10\pi(|\phi|-40)/180))/2.
\end{equation}
 This distribution gives $f=0$ at the equator and $f=1$ at the poles, with a cross-over latitude of 40$^\circ$. We then integrated the computed spectra over the disc to determine the disc-averaged geometric reflectivity in the various filters. Finally, to account for the fact that our reference reflectivity spectrum was calculated from disc-averaged data,  rather than equatorial data as assumed by \citet{irwin24}, and thus already partly includes the hood as seen in 2002, the final reflectivities were multiplied by a factor of 0.99 to ensure a good match to the observed Str\"omgren-$b$ and Str\"omgren-$y$ Lowell Observatory reflectivities of \cite{lockwood19}. Our simulated Str\"omgren reflectivity curves are compared with the observed Lowell reflectivities in  Fig. \ref{fig:seasonal_meas_sim} and also to HST/WFC3 observations from 2014 to 2022 \citep{irwin24}. This figure is very similar to Fig. 13 of \citet{irwin24} and shows moderately good agreement between the observed and simulated curves between 1985 and 2022. However, the observed reflectivities are lower than those simulated from 1950 to 1985 (especially at green wavelengths) for reasons that are currently unclear. \citet{irwin24} suggest that the reflectivity may be dependent on Uranus's mean distance from the Sun. Uranus was closest to the Sun in 1965 and furthest in 2009 and \citet{irwin24} suggest that increased solar flux may lead to a higher concentration of dark chromophore particles near perihelion, causing asymmetry in the reflectivity cycles. 
 
 The apparent correlation between giant planet atmospheric colouring agents, atmospheric dynamics and solar irradiation is under active investigation. \citet{hoyos09} purported that colour changes over Oval BA, one of Jupiter’s largest storms, might have been caused by increased insolation after the material was lofted to a higher altitude, in turn increasing the rate of photochemical reactions affecting the colour change. A correlation between chromophore darkness and UV fluence was also noted by \citet{carlson16}, who performed laboratory experiments to find a candidate for Jupiter's red chromophore \citep{carlson16} and may indicate that the deeper redness of Jupiter's Great Red Spot is caused by the haze particles being trapped in a vortex and exposed to sunlight for  extended periods. This correlation between UV irradiation and chromophore colour is currently under further investigation \citep{carlson23}.
 
\begin{figure*}
	\includegraphics[width=\textwidth]{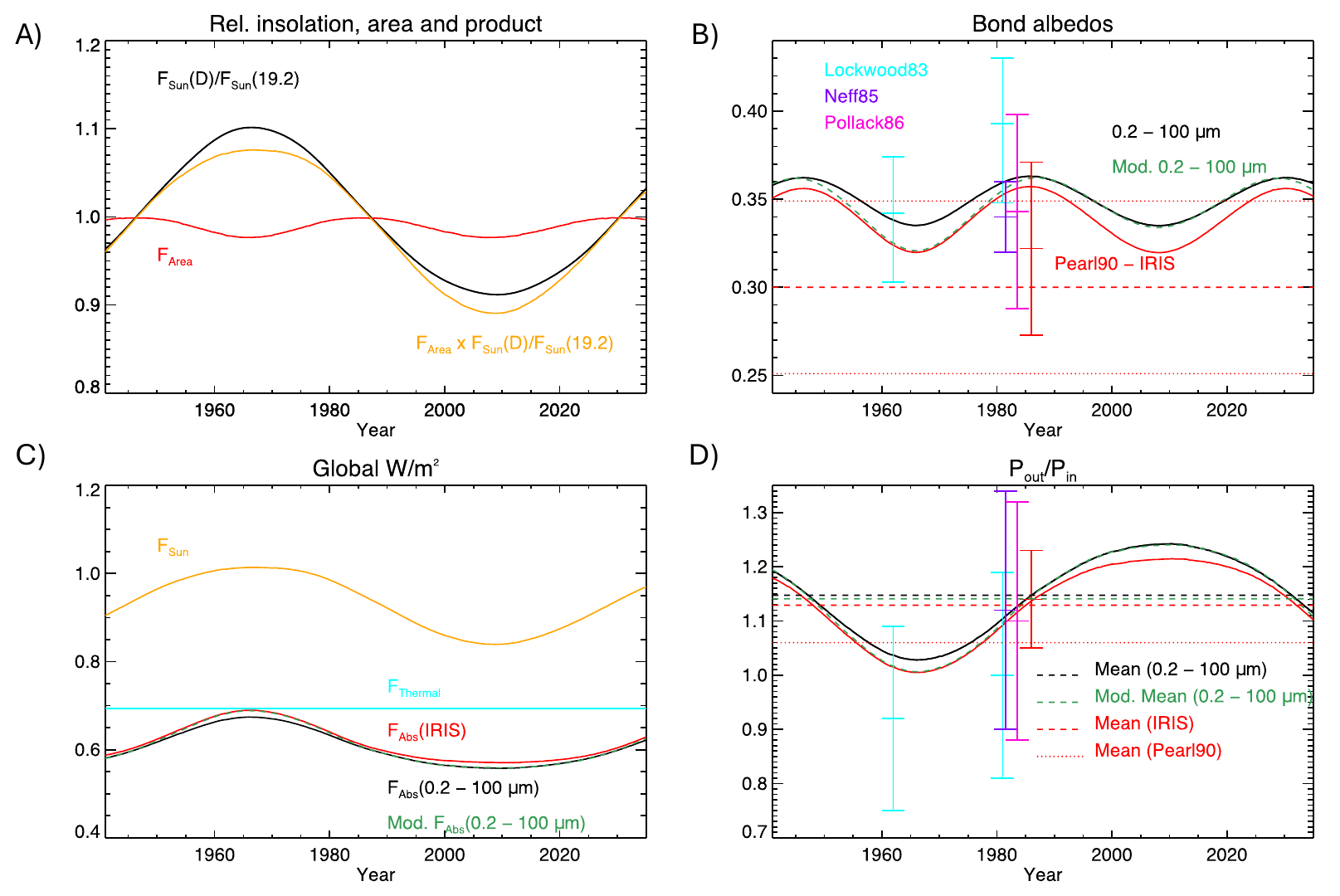}
    \caption{Calculations of Uranus's orbitally-averaged Bond albedo and heat budget. Panel A: Orbital variation of the incident solar flux at Uranus relative to the standard Uranus distance of 19.2 AU (black), compared with the projected area of Uranus relative to a sphere with Uranus's equatorial radius (red); the orange line is the product of the two. Panel B: Computed variation of Bond albedo with season, averaged over all wavelengths (0.2 -- 100 $\mu$m, black), or averaged over the Voyager/IRIS radiometer function (red). These variations are compared with estimates of the bolometric Bond albedos listed in Table \ref{tab:albedos} from \citet{lockwood83}, \citet{neff85}, \citet{pollack86}, and \citet{pearl90} and colour-coded as shown. The orbital-mean Bond albedo from \citet{pearl90} is shown as the dashed red line, with error limits shown as red dotted lines. Panel C: Seasonal energy budget of Uranus's atmosphere. The incident solar flux per unit area of Uranus's atmosphere is shown as orange, while the other lines show the solar fluxes absorbed by Uranus's atmosphere assuming Bond albedo variations calculated over all wavelengths (0.2 -- 100 $\mu$m, black line), or averaged over the Voyager/IRIS radiometer function (red line). These fluxes are compared with the assumed outgoing thermal flux of $0.693 \pm 0.013$ W m$^{-2}$ \citep[cyan line, ][]{pearl90}. Panel D: Seasonal variation of the $P_\mathrm{out}/P_\mathrm{in}$ flux ratio, averaged over all wavelengths (0.2 -- 100 $\mu$m, black), or over the Voyager/IRIS radiometer function (red). The orbital mean values are also shown as dashed lines and compared with the previous estimates listed in Table \ref{tab:albedos} and using the same colours as in Panel B. In panels B -- D, the dashed green line shows the 0.2 -- 100 $\mu$m calculations derived from the empirically adjusted $\Bar{p}_{(0.2-1.9 \mu \mathrm{m})}$ curve shown in Fig. \ref{fig:seasonal_meas_sim}.} 
    \label{fig:seasonal_calc}
\end{figure*}

 To see if the observed asymmetry in the seasonal curves might affect our orbitally-averaged heat balance calculations, we also generated an empirical fit to the Str\"omgren-$y$ curve using:
 \begin{equation}
    R^\prime(t)= \overline{R(t)} - 0.005 + 1.5(R(t)- \overline{R(t)})-0.1(S(t)-1)
\end{equation}
where $R(t)$ is our  simulated Str\"omgren-$y$ seasonal curve (shown as solid green in Fig. \ref{fig:seasonal_meas_sim}), $\overline{R(t)}$ is its orbital average, and $S(t) = F_\mathrm{Sun}(t)/F_\mathrm{Sun0}$, where $F_\mathrm{Sun}(t)$ is the solar insolation curve, shown later in Fig. \ref{fig:seasonal_calc}, and $F_\mathrm{Sun0}$ is the insolation when Uranus is at a distance of 19.2 AU from the Sun.

Figure \ref{fig:seasonal_meas_sim} also shows reflectivity curves calculated when averaged over the Voyager/ISS filters, the Voyager/IRIS radiometer response function, and finally when averaged over all calculation wavelengths (0.2 -- 1.9 $\mu$m). It can be seen that the amplitude of the reflectivity variations increases as we go to longer wavelengths, due to the fact that at longer wavelengths the polar hood becomes clearer and less masked by Rayleigh scattering. Averaging over all calculation wavelengths $\Bar{p}_{(0.2-1.9 \mu \mathrm{m})}$ can be seen to vary from roughly 0.27 to 0.29 over Uranus's year. We used these curves to estimate how much more reflective Uranus was in 1986 (during the Voyager-2 encounter) compared with the year of our HST/STIS observations in 2002, and phase integral curves including these factors are also shown in Fig. \ref{fig:phase_curves} as dashed lines. In particular, averaging over the Voyager/IRIS radiometer response function we estimate that Uranus should have been more reflective by 9.5\% in 1986 than it was in 2002. Hence, with this additional correction we estimate the likely geometric and Bond albedos in 1986, averaged over the IRIS response function to have been $\Bar{p}_\mathrm{IRIS}=0.256$ and $\Bar{A}_\mathrm{IRIS} = 0.346$, which are considerably larger than the estimates of \citet{pearl90}, but still within the estimated error of their values. Finally, Fig. \ref{fig:seasonal_meas_sim} also shows an additional 0.2 -- 1.9 $\mu$m curve derived by scaling the empirical Str\"omgren-$y$ seasonal curve by the mean ratio between the computed curves integreted over the Str\"omgren-$y$ filter and over the range 0.2 to 1.9 $\mu$m. This is used later to investigate how much the asymmetry in the observed Str\"omgren-$y$ seasonal curve might affect the orbitally-averaged heat balance.

Using the calculated seasonal curve for $\Bar{p}_{(0.2-1.9 \mu \mathrm{m})}$ we computed the seasonal cycle of the Bond albedo using $A = p q$, taking $\Bar{q}_{(0.2-1.9 \mu \mathrm{m})}$ from Table \ref{tab:calculations} to be 1.36. In the previous section we saw that the Bond albedo estimated from averaging from 0.2 -- 1.9 $\mu$m is likely lower than the total bolometric Bond albedo by a factor of 1.08 and hence our final bolometric albedo  $\Bar{A}_{(0.2-100 \mu \mathrm{m})}$ is corrected by this factor, assuming it does not change with season. However, to compute the total power absorbed by Uranus, we need to consider not only how the Bond albedo varies with season, but also that the incident solar flux varies during Uranus's eccentric orbit by approximately $\pm 10$\%, as can be seen in Fig. \ref{fig:seasonal_calc}A. In addition, the area of Uranus presented to the Sun varies since Uranus is significantly oblate and has a high axial inclination. Hence, the presented area  is maximum at the solstices and minimum at the equinoxes, varying by approximately $\pm$ 1\%. Combining these effects we can compute the total solar power intercepted by Uranus and divide this by the planet's total surface area to determine the average incident solar flux per unit area of the atmosphere, $F_\mathrm{Sun}$. We find that this varies from 0.84 to 1.01 W/m$^2$ during Uranus's orbit (Fig. \ref{fig:seasonal_calc}C).  Fig. \ref{fig:seasonal_calc}B shows how the Bond albedo varies with time, either averaging over all wavelengths (0.2 - 100 $\mu$m), or averaging over the Voyager/IRIS radiometer filter.  Combining these effects, we can compute the bolometric value of the absorbed flux  per unit area $F_\mathrm{abs}$, which we show in Fig. \ref{fig:seasonal_calc}C. The bolometric value of this  can be seen to vary from 0.558 to 0.674 W/m$^{2}$, with a mean value of 0.604 W/m$^{2}$. The confidence of this value depends on several factors: 
\begin{enumerate}
    \item The accuracy of the HST/STIS, Gemini/NIFS and IRTF/SpeX observations fitted to by \citet{irwin22}, which we estimated previously to be 2.8\%;
    
    \item The accuracy of our seasonal cycle model, which we show to be consistent with the Lowell observations of \citet{lockwood19}, but which we have extrapolated here initially to 1.9 $\mu$m in our radiative transfer calculation. The accuracy of this extrapolation is more difficult to evaluate, but since the bulk of Uranus's reflectivity comes from visible wavelengths, we believe that 2\% is a conservative estimate of the likely error;
    
    \item The accuracy of using the same phase integral value of $q=1.36$ for all seasons and all wavelengths. We model the poles to be more reflective to match the seasonal cycles, but we have no way of assessing how much the phase integral varies with season. Again, to be conservative we assume here 2\% for this uncertainty. 

    \item The accuracy of the factor of 1.08 we used to convert $\Bar{A}$ calculated from 0.2 -- 1.9 $\mu$m to all wavelengths (0.2 -- 100 $\mu$m). This was computed from our reference holistic model, which did not have a separate polar hood and hence may in reality vary with season. To account for this we again conservatively estimate this value to be accurate to 2\%.

    \item The observed asymmetry in the seasonal reflectivities reported by \citet{lockwood19}, but not reproduced in our simulations. To attempt to quantify this error, we also made calculations using the empirical $\Bar{p}_{(0.2-1.9 \mu \mathrm{m})}$ curve mentioned earlier and shown in Figs. \ref{fig:seasonal_meas_sim} and \ref{fig:seasonal_calc}, which we report below.
    
\end{enumerate}

All other elements in our calculation come from well-established solar spectra, or geometrical considerations, which we estimate to have errors insignificant with those listed above. Adding all these components in quadrature we arrive at a likely percentage error for $A^*$ and $P_\mathrm{in}$ of $\pm$4.5\%. Hence, including all factors we find orbital mean averages of $\overline{A^*} = 0.349 \pm 0.016$ and $P_\mathrm{in} = 0.604 \pm 0.027$ W/m$^2$. To determine the flux ratio $P_\mathrm{out}/P_\mathrm{in}$ we assume $P_\mathrm{out}$ to be $0.693 \pm 0.013$ W/m$^2$, as determined by \citet{pearl90} from Voyager/IRIS spectrometer observations, and we conclude an orbital mean of $P_\mathrm{out}/P_\mathrm{in}$ of $1.15 \pm 0.06$  (i.e., 4.9\%), varying during Uranus's year from 1.03 to 1.24. 

%The  1.07 to 1.29, with a mean value of 1.200. If we instead average over the IRIS filter function this ratio is reduced to $P_\mathrm{out}/P_\mathrm{in} = 1.129$. Both annual mean estimates are also shown in Fig. \ref{fig:seasonal_calc}, and both can be seen to be considerably greater than 1.0.

To see what effect the observed asymmetry of Uranus's reflectivity curves \citep{lockwood19} might have on these calculations, we repeated them using the empirically adjusted $\Bar{p}_{(0.2-1.9 \mu \mathrm{m})}$ curve shown in Fig. \ref{fig:seasonal_meas_sim} and find values of $\overline{A^*}=0.345 \pm 0.016$, $P_\mathrm{in} = 0.607 \pm 0.027$ W/m$^2$, and $P_\mathrm{out}/P_\mathrm{in} = 1.14 \pm 0.06$, which are well within the error limits of our formal calculations and show that the observed asymmetry in the observed Str\"omgren-$y$ seasonal curve likely does not significantly affect Uranus's heat balance.

Finally,  if we instead average over the IRIS filter function we find orbital-mean values of $\overline{A}_\mathrm{IRIS}=0.338 \pm 0.015$, $P_\mathrm{in} = 0.614 \pm 0.027$ W/m$^2$, and $P_\mathrm{out}/P_\mathrm{in}(\mathrm{IRIS}) = 1.13 \pm 0.06$. These values are within error of those calculated by averaging from 0.2 -- 100 $\mu$m, showing that the assumption made by \citet{pearl90} that $A_\mathrm{IRIS} = A^*$ is reasonable when averaged over a whole Uranus year. The values are consistent with those of \citet{pearl90} of $\overline{A}_\mathrm{IRIS}=0.300 \pm 0.049$ and $P_\mathrm{out}/P_\mathrm{in} = 1.06 \pm 0.08$, but have smaller error bars.
We have taken great care in our study to conservatively estimate our error sources, and we have also taken great care to estimate how the albedo of Uranus varies over the year and how varying orbital distance and projected cross-sectional area affects the absorbed solar power. Even with our conservative error assumptions, our much more detailed and robust calculation leads to error bars on $P_\mathrm{out}/P_\mathrm{in}$ that are smaller than those estimated by \citet{pearl90} and hence we conclude that our solution is better constrained and more robust.

%\subsection{Subsection}

%Spectroscopic 

%\subsubsection{sub-sub-section}

\section{Discussion}

We have concluded in this study that the bolometric (i.e., averaged over all wavelengths) value of $P_\mathrm{out}/P_\mathrm{in}$ for Uranus is significantly greater than 1.0 and thus that Uranus is not in thermal equilibrium with the Sun. Hence, Uranus appears to be much less of an outlier compared with the other planets than previously thought, although the Uranian value of $P_\mathrm{out}/P_\mathrm{in}$ is still significantly smaller than for the other three giant planets, for which values of $P_\mathrm{out}/P_\mathrm{in} \sim 2$ are generally accepted. 

Our findings of a significant source of internal heat for Uranus are more in line with the expectations of radiative-convective modelling studies. \citet{marley99} find an internal heat flow less than $P_\mathrm{int} = 0.06$ W m$^{-2}$ results in model tropospheres that are significantly cooler than those observed by the Voyager radio-occultation experiment \citep{lindal87}. Assuming $P_\mathrm{out}=0.693 \pm 0.013$ W m$^{-2}$, the value of $P_\mathrm{out}/P_\mathrm{in} = 1.06$ determined by \citet{pearl90} gives $P_\mathrm{int} = 0.04$ W m$^{-2}$, which would appear to be too small. However, using our new value of $P_\mathrm{out}/P_\mathrm{in} = 1.15$ gives $P_\mathrm{int} = 0.10$ W m$^{-2}$, which is more consistent with \citet{marley99}. More recently, \citet{markham21} look at how condensation effects may have affected the thermal evolution of Uranus and Neptune. Such effects can inhibit convection, but the apparent absence of an internal source of heat in Uranus defied explanation in their study. Most recently, \citet{milcareck24} have performed radiative-convective modelling of the stratospheres of Uranus and Neptune using the holistic aerosol model of \citet{irwin22}. Although this paper does not make estimates of internal heat flow, the value of the Bond Albedo derived for Uranus of 0.35 is very similar to what we have determined here of $0.349 \pm 0.016$, adding confidence to our estimate.

We find that Uranus is furthest from thermal equilibrium near the northern summer equinox period, a period which has just passed (equinox was in 2007), which may help to explain the increased storm cloud activity seen in Uranus's atmosphere near this time. It may be that increased solar heating at other times helps to stabilise the atmosphere by heating the upper levels and suppressing convection from the interior. This may thus help explain why Uranus appeared to be so quiescent and dormant during the Voyager 2 flyby in 1986. If this interpretation is correct, then it is likely that the planet's storm activity will now decrease as the incident solar flux increases and the upper levels of Uranus's atmosphere become warmer. In addition, as Uranus now starts to move closer to the Sun and the insolation starts to increase again, the Lowell observatory data of \citet{lockwood19} would suggest that Uranus should become slightly darker heading towards the next equinox in 2049 than it was in 2007. Continued observations are essential to confirm whether Uranus's seasonal albedo cycle really is asymmetric. %This asymmetry in the seasonal forcing may also explain the asymmetry seen in the record of   Str\"omgren filter-averaged magnitudes recorded at the Lowell observatory \citep{lockwood19}.

In addition to the seasonal cycles considered here, there is also a small variation in Uranus's albedo detected in the Lowell Observatory data of \citet{lockwood19} on the timescale of the solar cycle \citep[i.e. $\sim$11 years, ][]{aplin17}. There are two mechanisms suggested to account for this solar modulation: nucleation onto ions or electrons created by galactic cosmic rays (GCR),  or UV-induced aerosol colour changes, which are also suggested to affect the albedo of Neptune \citep[e.g.,][]{aplin16, roman22}.  Given that a slower seasonal variation of albedo is seen in the Lowell Observatory data \citep{lockwood19} that may be related to Uranus's distance from the Sun, this suggests that the second explanation may be more likely in this case. However, such solar-cycle variations are too rapid and too small to affect our orbital-mean estimates here.

Finally, our estimate for $P_\mathrm{out}/P_\mathrm{in}$  is critically dependent on the assumed value of the outgoing thermal flux, which we have assumed to be $P_\mathrm{out}=0.693 \pm 0.013$ W m$^{-2}$ \citep{pearl90} and to not vary with season. If, however, there has been a miscalibration of this value, or if the outgoing thermally emitted flux varies during the Uranian year then this will affect the flux ratio, as indeed would any unmodelled seasonal variations in the phase integral $q^*$, which we have here assumed to have a constant value of $q^* = 1.36 \pm 0.03$.

\section{Conclusions}

In this study we used the modified `holistic' aerosol model of \citet{irwin22}, \citet{james23}, and \citet{irwin24} to compute a new estimate of the bolometric Bond albedo of Uranus in the early 2000s. Then, using a seasonal model of Uranus based on \citet{irwin24}, we have calculated how this albedo varies with season and make a new estimate of the orbitally-averaged mean solar flux absorbed by Uranus, comparing this with the assumed outgoing thermal flux of $0.693 \pm 0.013$ W/m$^2$, determined from Voyager-2/IRIS observations by \citet{pearl90}. We arrive at the following conclusions.

\begin{enumerate}

\item  Averaging from 0.2 -- 100 $\mu$m, and using a fixed phase integral value of $q^* = 1.36$ at all seasons we find an orbital-average bolometric Bond albedo of $\overline{A^*} = 0.349 \pm 0.016$, noting that this varies significantly during the year from 0.335 to 0.363.

%\item Averaging over the Voyager/IRIS-radiometer responsivity function \citep{hanel81} we find an orbital average Bond albedo of $\overline{A}_\mathrm{IRIS}=0.338 \pm 0.015$, again noting that this varies significantly during the year from 0.319 to 0.357.

\item  Combining these computed reflectivity curves with solar flux and projected area orbital variations, and averaging from 0.2 -- 100 $\mu$m we find the total solar power absorbed by Uranus to be $0.604 \pm 0.027$ W/m$^2$, varying during the year from 0.558 to 0.674 W/m$^2$.

%\item  Combining these computed reflectivity curves with solar flux and projected area orbital variations, and averaging over the Voyager/IRIS-radiometer responsivity function we find the total solar power absorbed by Uranus to be $0.614 \pm 0.030$ W/m$^2$, varying during the year from 0.570 to 0.690 W/m$^2$.

\item Assuming the orbital mean outgoing thermal flux of $0.693 \pm 0.013$ W/m$^2$ from Voyager-2/IRIS \citep{pearl90}, we find, averaging from 0.2 -- 100 $\mu$m, that $P_\mathrm{out}/P_\mathrm{in} = 1.15 \pm 0.06$, varying during the year from 1.03 to 1.24. 

\item If we instead use the empirical geometric albedo curve derived from the observed asymmetric Str\"omgren-$y$ seasonal curve of \citet{lockwood19} and extrapolate to 0.2 -- 100 $\mu$m we find values of $\overline{A^*}=0.345 \pm 0.016$, $P_\mathrm{in} = 0.607 \pm 0.027$ W/m$^2$, and $P_\mathrm{out}/P_\mathrm{in} = 1.14 \pm 0.06$, which are well within the error limits of our formal calculation. Hence, we conclude that the observed asymmetry in the observed Str\"omgren-$y$ seasonal curve does not significantly affect Uranus's heat balance.

\item Finally, if we instead average over the Voyager/IRIS-radiometer responsivity function \citep{hanel81}, we find values of $\overline{A}_\mathrm{IRIS}=0.338 \pm 0.015$, $P_\mathrm{in} = 0.614 \pm 0.027$ W/m$^2$, and $P_\mathrm{out}/P_\mathrm{in} = 1.13 \pm 0.06$. These are just within the error limits of our 0.2 -- 100 $\mu$m calculations and show that assuming $A_\mathrm{IRIS} = A^*$, as was done by \citet{pearl90}, is a reasonable approximation.
%\item Assuming the same orbital mean outgoing thermal flux and averaging over the Voyager/IRIS-radiometer responsivity function, we find that $P_\mathrm{out}/P_\mathrm{in} = 1.13 \pm 0.06$, varying during the year from 1.00 to 1.22.

\end{enumerate}

While we have used our simulations to greatly improve the estimated values of  $\overline{A^*}$ and $P_\mathrm{in}$ for Uranus, our estimate of $P_\mathrm{out}/P_\mathrm{in}$ is critically dependent on the assumed value of the outgoing thermal flux $P_\mathrm{out}=0.693 \pm 0.013$ W m$^{-2}$ \citep{pearl90}. We have assumed this value to be correct and have also assumed that it does not vary with time. If, however, there has been a miscalibration of this value, or if, in fact, the outgoing thermally emitted flux  varies with season, then this will affect the flux ratio. We thus recommend that the value of the outgoing thermal flux from Uranus's atmosphere be restudied, perhaps using more recent ground-based and JWST observations. In the meantime, assuming $P_\mathrm{out}=0.693 \pm 0.013$ W m$^{-2}$
 we conclude that although Uranus is not in thermal equilibrium with the incident solar flux, $P_\mathrm{out}/P_\mathrm{in}$ is still significantly less than for the other three giant planets, for which values of $P_\mathrm{out}/P_\mathrm{in} \sim 2$ are generally accepted.  We also note that the high eccentricity of Uranus's orbit leads to significant change in the incident solar flux, which if correlated with the production of dark photochemical haze products may help to explain the asymmetry in Uranus's seasonal albedo cycle observed in Lowell Observatory observations \citep{lockwood19}. To explore this further, the next step in this project will be to analyze Hubble visible wavelength data from 1994 through 2025 to gain a clearer understanding of how Uranus's reflectivity changes at other wavelengths while the planet moved from just before the northern spring equinox towards northern summer solstice. 

\section*{Acknowledgements}

We are grateful to the United Kingdom Science and Technology Facilities Council for funding the early stages of this research (Irwin: ST/S000461/1). This review includes observations made with the NASA/ESA Hubble Space Telescope obtained from the Space Telescope Science Institute, which is operated by the Association of Universities for Research in Astronomy, Inc. (AURA), under NASA contract NAS 5–26555. These observations are associated with STIS programme(s) GO9035, GO9330, GO12894, GO14113, and WFC3 programmes GO13937, GO14334, GO14756, GO15262, GO15502, GO15929, GO16266, GO16790, GO16995. AAS was supported by a grant associated with continuing program GO13937.

%%%%%%%%%%%%%%%%%%%%%%%%%%%%%%%%%%%%%%%%%%%%%%%%%%
\section*{Data Availability}

The data underlying this article will be shared on reasonable request to the corresponding author. %Please note, however, that the Hubble data are publicly available at the Mikulski Archive for Space Telescopes (\url{https://mast.stsci.edu/}) under the program numbers listed in the acknowledgements. 

%(see \url{https://academic.oup.com/pages/open-research/research-data#Data%20Availability%20Statements}.

%The inclusion of a Data Availability Statement is a requirement for articles published in MNRAS. Data Availability Statements provide a standardised format for readers to understand the availability of data underlying the research results described in the article. The statement may refer to original data generated in the course of the study or to third-party data analysed in the article. The statement should describe and provide means of access, where possible, by linking to the data or providing the required accession numbers for the relevant databases or DOIs.



\bibliographystyle{mnras}
\bibliography{bibliography} % if your bibtex file is called example.bib

\begin{thebibliography}{}
\makeatletter
\relax
\def\mn@urlcharsother{\let\do\@makeother \do\$\do\&\do\#\do\^\do\_\do\%\do\~}
\def\mn@doi{\begingroup\mn@urlcharsother \@ifnextchar [ {\mn@doi@}
  {\mn@doi@[]}}
\def\mn@doi@[#1]#2{\def\@tempa{#1}\ifx\@tempa\@empty \href
  {http://dx.doi.org/#2} {doi:#2}\else \href {http://dx.doi.org/#2} {#1}\fi
  \endgroup}
\def\mn@eprint#1#2{\mn@eprint@#1:#2::\@nil}
\def\mn@eprint@arXiv#1{\href {http://arxiv.org/abs/#1} {{\tt arXiv:#1}}}
\def\mn@eprint@dblp#1{\href {http://dblp.uni-trier.de/rec/bibtex/#1.xml}
  {dblp:#1}}
\def\mn@eprint@#1:#2:#3:#4\@nil{\def\@tempa {#1}\def\@tempb {#2}\def\@tempc
  {#3}\ifx \@tempc \@empty \let \@tempc \@tempb \let \@tempb \@tempa \fi \ifx
  \@tempb \@empty \def\@tempb {arXiv}\fi \@ifundefined
  {mn@eprint@\@tempb}{\@tempb:\@tempc}{\expandafter \expandafter \csname
  mn@eprint@\@tempb\endcsname \expandafter{\@tempc}}}

\bibitem[\protect\citeauthoryear{{Aplin} \& {Harrison}}{{Aplin} \&
  {Harrison}}{2016}]{aplin16}
{Aplin} K.~L.,  {Harrison} R.~G.,  2016, \mn@doi [Nature Communications]
  {10.1038/ncomms11976}, \href
  {https://ui.adsabs.harvard.edu/abs/2016NatCo...711976A} {7, 11976}

\bibitem[\protect\citeauthoryear{{Aplin} \& {Harrison}}{{Aplin} \&
  {Harrison}}{2017}]{aplin17}
{Aplin} K.~L.,  {Harrison} R.~G.,  2017, \mn@doi [\grl] {10.1002/2017GL075374},
  \href {https://ui.adsabs.harvard.edu/abs/2017GeoRL..4412083A} {44, 12,083}

\bibitem[\protect\citeauthoryear{{Caldwell}, {Owen}, {Rivolo}, {Moore}, {Hunt}
  \& {Butterworth}}{{Caldwell} et~al.}{1981}]{caldwell81}
{Caldwell} J.,  {Owen} T.,  {Rivolo} A.~R.,  {Moore} V.,  {Hunt} G.~E.,
  {Butterworth} P.~S.,  1981, \mn@doi [\aj] {10.1086/112888}, \href
  {https://ui.adsabs.harvard.edu/abs/1981AJ.....86..298C} {86, 298}

\bibitem[\protect\citeauthoryear{{Carlson}, {Baines}, {Anderson}, {Filacchione}
   \& {Simon}}{{Carlson} et~al.}{2016}]{carlson16}
{Carlson} R.~W.,  {Baines} K.~H.,  {Anderson} M.~S.,  {Filacchione} G.,
  {Simon} A.~A.,  2016, \mn@doi [\icarus] {10.1016/j.icarus.2016.03.008}, \href
  {https://ui.adsabs.harvard.edu/abs/2016Icar..274..106C} {274, 106}

\bibitem[\protect\citeauthoryear{{Carlson}, {Dahl}, {Baines}, {Jovanovic}  \&
  {Anderson}}{{Carlson} et~al.}{2023}]{carlson23}
{Carlson} R.,  {Dahl} E.,  {Baines} K.,  {Jovanovic} L.,   {Anderson} M.,
  2023, in AAS/Division for Planetary Sciences Meeting Abstracts \#55. p.
  505.02

\bibitem[\protect\citeauthoryear{{Chance} \& {Kurucz}}{{Chance} \&
  {Kurucz}}{2010}]{chance10}
{Chance} K.,  {Kurucz} R.~L.,  2010, \jqsrt, 111, 1289

\bibitem[\protect\citeauthoryear{{Courtin}, {Lena}, {de Muizon}, {Rouan},
  {Nicollier}  \& {Wijnbergen}}{{Courtin} et~al.}{1979}]{courtin79}
{Courtin} R.,  {Lena} P.,  {de Muizon} M.,  {Rouan} D.,  {Nicollier} C.,
  {Wijnbergen} J.,  1979, \mn@doi [\icarus] {10.1016/0019-1035(79)90196-9},
  \href {https://ui.adsabs.harvard.edu/abs/1979Icar...38..411C} {38, 411}

\bibitem[\protect\citeauthoryear{{Crawford} \& {Barnes}}{{Crawford} \&
  {Barnes}}{1970}]{crawford70}
{Crawford} D.~L.,  {Barnes} J.~V.,  1970, \mn@doi [\aj] {10.1086/111051}, \href
  {https://ui.adsabs.harvard.edu/abs/1970AJ.....75..978C} {75, 978}

\bibitem[\protect\citeauthoryear{{Danielson}, {Kupferman}, {Johnson}  \&
  {Soderblom}}{{Danielson} et~al.}{1981}]{danielson81}
{Danielson} G.~E.,  {Kupferman} P.~N.,  {Johnson} T.~V.,   {Soderblom} L.~A.,
  1981, \mn@doi [\jgr] {10.1029/JA086iA10p08683}, \href
  {https://ui.adsabs.harvard.edu/abs/1981JGR....86.8683D} {86, 8683}

\bibitem[\protect\citeauthoryear{{Fletcher}, {de Pater}, {Orton}, {Hofstadter},
  {Irwin}, {Roman}  \& {Toledo}}{{Fletcher} et~al.}{2020}]{fletcher20}
{Fletcher} L.~N.,  {de Pater} I.,  {Orton} G.~S.,  {Hofstadter} M.~D.,  {Irwin}
  P. G.~J.,  {Roman} M.~T.,   {Toledo} D.,  2020, \mn@doi [\ssr]
  {10.1007/s11214-020-00646-1}, \href
  {https://ui.adsabs.harvard.edu/abs/2020SSRv..216...21F} {216, 21}

\bibitem[\protect\citeauthoryear{{Hanel}, {Conrath}, {Herath}, {Kunde}  \&
  {Pirraglia}}{{Hanel} et~al.}{1981}]{hanel81}
{Hanel} R.,  {Conrath} B.,  {Herath} L.,  {Kunde} V.,   {Pirraglia} J.,  1981,
  \mn@doi [\jgr] {10.1029/JA086iA10p08705}, \href
  {https://ui.adsabs.harvard.edu/abs/1981JGR....86.8705H} {86, 8705}

\bibitem[\protect\citeauthoryear{{Hanel}, {Conrath}, {Kunde}, {Pearl}  \&
  {Pirraglia}}{{Hanel} et~al.}{1983}]{hanel83}
{Hanel} R.~A.,  {Conrath} B.~J.,  {Kunde} V.~G.,  {Pearl} J.~C.,   {Pirraglia}
  J.~A.,  1983, \mn@doi [\icarus] {10.1016/0019-1035(83)90147-1}, \href
  {https://ui.adsabs.harvard.edu/abs/1983Icar...53..262H} {53, 262}

\bibitem[\protect\citeauthoryear{{Hildebrand}, {Loewenstein}, {Harper},
  {Orton}, {Keene}  \& {Whitcomb}}{{Hildebrand} et~al.}{1985}]{hildebrand85}
{Hildebrand} R.~H.,  {Loewenstein} R.~F.,  {Harper} D.~A.,  {Orton} G.~S.,
  {Keene} J.,   {Whitcomb} S.~E.,  1985, \mn@doi [\icarus]
  {10.1016/0019-1035(85)90039-9}, \href
  {https://ui.adsabs.harvard.edu/abs/1985Icar...64...64H} {64, 64}

\bibitem[\protect\citeauthoryear{{Hueso}, {Guillot}  \&
  {S{\'a}nchez-Lavega}}{{Hueso} et~al.}{2020}]{hueso20}
{Hueso} R.,  {Guillot} T.,   {S{\'a}nchez-Lavega} A.,  2020, \mn@doi
  [Philosophical Transactions of the Royal Society of London Series A]
  {10.1098/rsta.2019.0476}, \href
  {https://ui.adsabs.harvard.edu/abs/2020RSPTA.37890476H} {378, 20190476}

\bibitem[\protect\citeauthoryear{{Irwin} et~al.,}{{Irwin}
  et~al.}{2008}]{irwin08}
{Irwin} P.~G.~J.,  et~al., 2008, \mn@doi [\jqsrt]
  {10.1016/j.jqsrt.2007.11.006}, \href
  {https://ui.adsabs.harvard.edu/abs/2008JQSRT.109.1136I} {109, 1136}

\bibitem[\protect\citeauthoryear{{Irwin}, {Teanby}, {Davis}, {Fletcher},
  {Orton}, {Tice}  \& {Kyffin}}{{Irwin} et~al.}{2011}]{irwin11}
{Irwin} P.~G.~J.,  {Teanby} N.~A.,  {Davis} G.~R.,  {Fletcher} L.~N.,  {Orton}
  G.~S.,  {Tice} D.,   {Kyffin} A.,  2011, \mn@doi [\icarus]
  {10.1016/j.icarus.2010.12.018}, \href
  {https://ui.adsabs.harvard.edu/abs/2011Icar..212..339I} {212, 339}

\bibitem[\protect\citeauthoryear{{Irwin} et~al.,}{{Irwin}
  et~al.}{2022}]{irwin22}
{Irwin} P.~G.~J.,  et~al., 2022, \mn@doi [\jgr:Planets] {10.1029/2022JE007189},
  \href {https://ui.adsabs.harvard.edu/abs/2022JGRE..12707189I} {127, e07189}

\bibitem[\protect\citeauthoryear{{Irwin} et~al.,}{{Irwin}
  et~al.}{2024}]{irwin24}
{Irwin} P. G.~J.,  et~al., 2024, \mn@doi [\mnras] {10.1093/mnras/stad3761},
  \href {https://ui.adsabs.harvard.edu/abs/2024MNRAS.527.11521} {527, 11521}

\bibitem[\protect\citeauthoryear{{James} et~al.,}{{James}
  et~al.}{2023}]{james23}
{James} A.,  et~al., 2023, \mn@doi [\jgr:Planets] {10.1029/2023JE007904}, \href
  {https://ui.adsabs.harvard.edu/abs/2023JGRE..12807904J} {128, e07904}

\bibitem[\protect\citeauthoryear{{Karkoschka} \& {Tomasko}}{{Karkoschka} \&
  {Tomasko}}{2009}]{kark09}
{Karkoschka} E.,  {Tomasko} M.,  2009, \mn@doi [Icarus]
  {10.1016/j.icarus.2009.02.010}, \href
  {https://ui.adsabs.harvard.edu/abs/2009Icar..202..287K} {202, 287}

\bibitem[\protect\citeauthoryear{{Kurucz}}{{Kurucz}}{1993}]{kurucz93}
{Kurucz} R.,  1993, ATLAS9 Stellar Atmosphere Programs and 2 km/s grid. Kurucz
  CD-ROM No. 13. Cambridge, \href
  {https://ui.adsabs.harvard.edu/abs/1993KurCD..13.....K} {13}

\bibitem[\protect\citeauthoryear{{Li} \& {Ingersoll}}{{Li} \&
  {Ingersoll}}{2015}]{li15}
{Li} C.,  {Ingersoll} A.~P.,  2015, \mn@doi [Nature Geoscience]
  {10.1038/ngeo2405}, \href
  {https://ui.adsabs.harvard.edu/abs/2015NatGe...8..398L} {8, 398}

\bibitem[\protect\citeauthoryear{{Li} et~al.,}{{Li} et~al.}{2018}]{li18}
{Li} L.,  et~al., 2018, \mn@doi [Nature Communications]
  {10.1038/s41467-018-06107-2}, \href
  {https://ui.adsabs.harvard.edu/abs/2018NatCo...9.3709L} {9, 3709}

\bibitem[\protect\citeauthoryear{{Lindal}, {Lyons}, {Sweetnam}, {Eshleman},
  {Hinson}  \& {Tyler}}{{Lindal} et~al.}{1987}]{lindal87}
{Lindal} G.~F.,  {Lyons} J.~R.,  {Sweetnam} D.~N.,  {Eshleman} V.~R.,  {Hinson}
  D.~P.,   {Tyler} G.~L.,  1987, \mn@doi [\jgr] {10.1029/JA092iA13p14987},
  \href {https://ui.adsabs.harvard.edu/abs/1987JGR....9214987L} {92, 14987}

\bibitem[\protect\citeauthoryear{{Lockwood}}{{Lockwood}}{2019}]{lockwood19}
{Lockwood} G.~W.,  2019, \mn@doi [\icarus] {10.1016/j.icarus.2019.01.024},
  \href {https://ui.adsabs.harvard.edu/abs/2019Icar..324...77L} {324, 77}

\bibitem[\protect\citeauthoryear{{Lockwood}, {Lutz}, {Thompson}  \&
  {Warnock}}{{Lockwood} et~al.}{1983}]{lockwood83}
{Lockwood} G.~W.,  {Lutz} B.~L.,  {Thompson} D.~T.,   {Warnock} A.,  1983,
  \mn@doi [\apj] {10.1086/160788}, \href
  {https://ui.adsabs.harvard.edu/abs/1983ApJ...266..402L} {266, 402}

\bibitem[\protect\citeauthoryear{{Markham} \& {Stevenson}}{{Markham} \&
  {Stevenson}}{2021}]{markham21}
{Markham} S.,  {Stevenson} D.,  2021, \mn@doi [Planet. Sci. J.]
  {10.3847/PSJ/ac091d}, \href
  {https://ui.adsabs.harvard.edu/abs/2021PSJ.....2..146M} {2, 146}

\bibitem[\protect\citeauthoryear{{Marley} \& {McKay}}{{Marley} \&
  {McKay}}{1999}]{marley99}
{Marley} M.~S.,  {McKay} C.~P.,  1999, \mn@doi [\icarus]
  {10.1006/icar.1998.6071}, \href
  {https://ui.adsabs.harvard.edu/abs/1999Icar..138..268M} {138, 268}

\bibitem[\protect\citeauthoryear{{Milcareck} et~al.,}{{Milcareck}
  et~al.}{2024}]{milcareck24}
{Milcareck} G.,  et~al., 2024, \mn@doi [\aap] {10.1051/0004-6361/202348987},
  \href {https://ui.adsabs.harvard.edu/abs/2024A&A...686A.303M} {686, A303}

\bibitem[\protect\citeauthoryear{{Minnaert}}{{Minnaert}}{1941}]{minnaert41}
{Minnaert} M.,  1941, \mn@doi [\apj] {10.1086/144279}, \href
  {https://ui.adsabs.harvard.edu/abs/1941ApJ....93..403M} {93, 403}

\bibitem[\protect\citeauthoryear{{Neff}, {Humm}, {Bergstralh}, {Cochran},
  {Cochran}, {Barker}  \& {Tull}}{{Neff} et~al.}{1984}]{neff84}
{Neff} J.~S.,  {Humm} D.~C.,  {Bergstralh} J.~T.,  {Cochran} A.~L.,  {Cochran}
  W.~D.,  {Barker} E.~S.,   {Tull} R.~G.,  1984, \mn@doi [\icarus]
  {10.1016/0019-1035(84)90186-6}, \href
  {https://ui.adsabs.harvard.edu/abs/1984Icar...60..221N} {60, 221}

\bibitem[\protect\citeauthoryear{{Neff}, {Ellis}, {Apt}  \&
  {Bergstralh}}{{Neff} et~al.}{1985}]{neff85}
{Neff} J.~S.,  {Ellis} T.~A.,  {Apt} J.,   {Bergstralh} J.~T.,  1985, \mn@doi
  [\icarus] {10.1016/0019-1035(85)90185-X}, \href
  {https://ui.adsabs.harvard.edu/abs/1985Icar...62..425N} {62, 425}

\bibitem[\protect\citeauthoryear{{Nettelmann}, {Wang}, {Fortney}, {Hamel},
  {Yellamilli}, {Bethkenhagen}  \& {Redmer}}{{Nettelmann}
  et~al.}{2016}]{nettelmann16}
{Nettelmann} N.,  {Wang} K.,  {Fortney} J.~J.,  {Hamel} S.,  {Yellamilli} S.,
  {Bethkenhagen} M.,   {Redmer} R.,  2016, \mn@doi [Icarus]
  {10.1016/j.icarus.2016.04.008}, \href
  {https://ui.adsabs.harvard.edu/abs/2016Icar..275..107N} {275, 107}

\bibitem[\protect\citeauthoryear{{Orton}}{{Orton}}{1985}]{orton85}
{Orton} G.~S.,  1985, in Bulletin of the American Astronomical Society. p.~745

\bibitem[\protect\citeauthoryear{{Pearl} \& {Conrath}}{{Pearl} \&
  {Conrath}}{1991}]{pearl91}
{Pearl} J.~C.,  {Conrath} B.~J.,  1991, \mn@doi [\jgr] {10.1029/91JA01087},
  \href {https://ui.adsabs.harvard.edu/abs/1991JGR....9618921P} {96, 18921}

\bibitem[\protect\citeauthoryear{{Pearl}, {Conrath}, {Hanel}, {Pirraglia}  \&
  {Coustenis}}{{Pearl} et~al.}{1990}]{pearl90}
{Pearl} J.~C.,  {Conrath} B.~J.,  {Hanel} R.~A.,  {Pirraglia} J.~A.,
  {Coustenis} A.,  1990, \mn@doi [\icarus] {10.1016/0019-1035(90)90155-3},
  \href {https://ui.adsabs.harvard.edu/abs/1990Icar...84...12P} {84, 12}

\bibitem[\protect\citeauthoryear{{P{\'e}rez-Hoyos}, {S{\'a}nchez-Lavega},
  {Hueso}, {Garc{\'\i}a-Melendo}  \& {Legarreta}}{{P{\'e}rez-Hoyos}
  et~al.}{2009}]{hoyos09}
{P{\'e}rez-Hoyos} S.,  {S{\'a}nchez-Lavega} A.,  {Hueso} R.,
  {Garc{\'\i}a-Melendo} E.,   {Legarreta} J.,  2009, \mn@doi [\icarus]
  {10.1016/j.icarus.2009.06.024}, \href
  {https://ui.adsabs.harvard.edu/abs/2009Icar..203..516P} {203, 516}

\bibitem[\protect\citeauthoryear{{Plass}, {Kattawar}  \& {Catchings}}{{Plass}
  et~al.}{1973}]{plass73}
{Plass} G.~N.,  {Kattawar} G.~W.,   {Catchings} F.~E.,  1973, \mn@doi [\ao]
  {10.1364/AO.12.000314}, \href
  {https://ui.adsabs.harvard.edu/abs/1973ApOpt..12..314P} {12, 314}

\bibitem[\protect\citeauthoryear{{Pollack}, {Rages}, {Baines}, {Bergstralh},
  {Wenkert}  \& {Danielson}}{{Pollack} et~al.}{1986}]{pollack86}
{Pollack} J.~B.,  {Rages} K.,  {Baines} K.~H.,  {Bergstralh} J.~T.,  {Wenkert}
  D.,   {Danielson} G.~E.,  1986, \mn@doi [\icarus]
  {10.1016/0019-1035(86)90147-8}, \href
  {https://ui.adsabs.harvard.edu/abs/1986Icar...65..442P} {65, 442}

\bibitem[\protect\citeauthoryear{{Rayner}, {Cushing}  \& {Vacca}}{{Rayner}
  et~al.}{2009}]{rayner09}
{Rayner} J.~T.,  {Cushing} M.~C.,   {Vacca} W.~D.,  2009, \mn@doi [\apjs]
  {10.1088/0067-0049/185/2/289}, \href
  {https://ui.adsabs.harvard.edu/abs/2009ApJS..185..289R} {185, 289}

\bibitem[\protect\citeauthoryear{{Reinhardt}, {Chau}, {Stadel}  \&
  {Helled}}{{Reinhardt} et~al.}{2020}]{reinhardt20}
{Reinhardt} C.,  {Chau} A.,  {Stadel} J.,   {Helled} R.,  2020, \mn@doi
  [\mnras] {10.1093/mnras/stz3271}, \href
  {https://ui.adsabs.harvard.edu/abs/2020MNRAS.492.5336R} {492, 5336}

\bibitem[\protect\citeauthoryear{{Roman} et~al.,}{{Roman}
  et~al.}{2022}]{roman22}
{Roman} M.~T.,  et~al., 2022, \mn@doi [Planet. Sci. J.] {10.3847/PSJ/ac5aa4},
  \href {https://ui.adsabs.harvard.edu/abs/2022PSJ.....3...78R} {3, 78}

\bibitem[\protect\citeauthoryear{{Scheibe}, {Nettelmann}  \&
  {Redmer}}{{Scheibe} et~al.}{2021}]{scheibe21}
{Scheibe} L.,  {Nettelmann} N.,   {Redmer} R.,  2021, \mn@doi [Astronomy \&
  Astrophysics] {10.1051/0004-6361/202140663}, \href
  {https://ui.adsabs.harvard.edu/abs/2021A&A...650A.200S} {650, A200}

\bibitem[\protect\citeauthoryear{Sromovsky \& Fry}{Sromovsky \&
  Fry}{2005}]{sromovsky05}
Sromovsky L.,  Fry P.,  2005, \mn@doi [Icarus]
  {https://doi.org/10.1016/j.icarus.2005.07.022}, 179, 459

\bibitem[\protect\citeauthoryear{{Sromovsky}, {Karkoschka}, {Fry}, {Hammel},
  {de Pater}  \& {Rages}}{{Sromovsky} et~al.}{2014}]{sromovsky14}
{Sromovsky} L.~A.,  {Karkoschka} E.,  {Fry} P.~M.,  {Hammel} H.~B.,  {de Pater}
  I.,   {Rages} K.,  2014, \mn@doi [\icarus] {10.1016/j.icarus.2014.05.016},
  \href {https://ui.adsabs.harvard.edu/abs/2014Icar..238..137S} {238, 137}

\bibitem[\protect\citeauthoryear{{Sromovsky}, {Karkoschka}, {Fry}, {de Pater}
  \& {Hammel}}{{Sromovsky} et~al.}{2019}]{sromovsky19}
{Sromovsky} L.~A.,  {Karkoschka} E.,  {Fry} P.~M.,  {de Pater} I.,   {Hammel}
  H.~B.,  2019, \mn@doi [\icarus] {10.1016/j.icarus.2018.06.026}, \href
  {https://ui.adsabs.harvard.edu/abs/2019Icar..317..266S} {317, 266}

\bibitem[\protect\citeauthoryear{{Stevenson}}{{Stevenson}}{1986}]{stevenson86}
{Stevenson} D.~J.,  1986, in Lunar and Planetary Science Conference. pp
  1011--1012

\bibitem[\protect\citeauthoryear{{Stockman}}{{Stockman}}{2019}]{stockman19}
{Stockman} A.,  2019, \mn@doi [Current Opinion in Behavioral Sciences]
  {10.1016/j.cobeha.2019.06.005}, 30, 87

\bibitem[\protect\citeauthoryear{{Stockman} \& {Sharpe}}{{Stockman} \&
  {Sharpe}}{2000}]{stockman00}
{Stockman} A.,  {Sharpe} L.~T.,  2000, \mn@doi [Vision Research]
  {10.1016/S0042-6989(00)00021-3}, 40, 1711

\bibitem[\protect\citeauthoryear{{Vazan} \& {Helled}}{{Vazan} \&
  {Helled}}{2020}]{vazan20}
{Vazan} A.,  {Helled} R.,  2020, \mn@doi [Astronomy \& Astrophysics]
  {10.1051/0004-6361/201936588}, \href
  {https://ui.adsabs.harvard.edu/abs/2020A&A...633A..50V} {633, A50}

\bibitem[\protect\citeauthoryear{{Wang} et~al.,}{{Wang} et~al.}{2024}]{wang24}
{Wang} X.,  et~al., 2024, \mn@doi [Nature Communications]
  {10.1038/s41467-024-48969-9}, \href
  {https://ui.adsabs.harvard.edu/abs/2024NatCo..15.5045W} {15, 5045}

\bibitem[\protect\citeauthoryear{{Wenkert}}{{Wenkert}}{2023}]{wenkert23a}
{Wenkert} D.,  2023, {Restoring and Archiving Voyager 1 Cruise Images of Uranus
  and Neptune (RAV1CIUN) PDART Bundle}, \mn@doi{10.17189/T2R8-RK88}

\bibitem[\protect\citeauthoryear{{Wenkert}, {Kenyon}, {Kleinboehl}, {Dahl}  \&
  {Orton}}{{Wenkert} et~al.}{2023}]{wenkert23}
{Wenkert} D.~D.,  {Kenyon} M.~E.,  {Kleinboehl} A.,  {Dahl} E.,   {Orton}
  G.~S.,  2023, in Uranus Flagship: Investigations and Instruments for
  Cross-Discipline Science Workshop. p.~8048

\bibitem[\protect\citeauthoryear{{Younkin}}{{Younkin}}{1970}]{younkin70}
{Younkin} R.~L.,  1970, PhD thesis, University of California, Los Angeles

\makeatother
\end{thebibliography}

%\begin{thebibliography}{99}
%\end{thebibliography}

%%%%%%%%%%%%%%%%%%%%%%%%%%%%%%%%%%%%%%%%%%%%%%%%%%

%%%%%%%%%%%%%%%%% APPENDICES %%%%%%%%%%%%%%%%%%%%%

%\appendix
%\section{Title}
%\label{appendix:1}

%The response 

%%%%%%%%%%%%%%%%%%%%%%%%%%%%%%%%%%%%%%%%%%%%%%%%%%

% Don't change these lines
\bsp	% typesetting comment
\label{lastpage}
\end{document}